\documentclass[a4paper,11pt]{article}
\usepackage{jinstpub} % for details on the use of the package, please
\usepackage{graphics}
\usepackage{epsfig}
\usepackage{epstopdf}
\usepackage{subfigure}
\usepackage{upgreek}

\usepackage{amssymb}
\usepackage{amsthm}
\usepackage{amsmath}
\usepackage{breakurl}
\usepackage{multirow}
\usepackage{booktabs}
\usepackage{bm}

\usepackage{lineno}

\usepackage{colortbl}

\usepackage{xcolor}
\title{Optical response of highly reflective film used in the water Cherenkov muon veto of the XENON1T dark matter experiment}

\author[a,1]{Ch.~Geis,\note{Corresponding authors}}
\author[a]{C.~Grignon,}
\author[a,b,1]{U.~Oberlack,}
\author[a,b]{D.~Ram\'{i}rez Garc\'{i}a,}
\author[b]{Q.~Weitzel}

\affiliation[a]{Institute of Physics \& PRISMA Cluster of Excellence, Johannes Gutenberg University Mainz,\\ 55099 Mainz, Germany}
\affiliation[b]{PRISMA Detector Laboratory, Johannes Gutenberg University Mainz,\\55099 Mainz, Germany}

\emailAdd{geisch@uni-mainz.de, oberlack@uni-mainz.de}

\abstract{The XENON1T experiment is the most recent stage of the XENON Dark Matter Search, aiming for the direct detection of dark matter candidates, such as the Weakly Interacting Massive Particles (WIMPs). The projected sensitivity for the spin-independent WIMP-nucleon elastic scattering cross-section is $\sigma \approx 2 \times 10^{-47}\,\mathrm{cm}^2$ for a WIMP mass of $m_{\chi} = 50\,\mathrm{GeV/c^2}$. To reach its projected sensitivity, the background has to be reduced by two orders of magnitude compared to its predecessor XENON100. This requires a water Cherenkov muon veto surrounding the XENON1T TPC, both to shield external backgrounds and to tag muon-induced energetic neutrons through detection of a passing muon or the secondary shower induced by a muon interacting in the surrounding rock. The muon veto is instrumented with $84$ $8"$ PMTs with high quantum efficiency (QE) in the Cherenkov regime and the walls of the watertank are clad with the highly reflective DF2000MA foil by 3M. Here, we present a study of the reflective properties of this foil, as well as the measurement of its wavelength shifting (WLS) properties. Furthermore, we present the impact of reflectance and WLS on the detection efficiency of the muon veto, through the use of a Monte Carlo simulation carried out with the Geant4 toolkit. The measurements yield a specular reflectance of $\approx100\%$ for wavelengths larger than $400\,$nm, while $\approx90\%$ of the incoming light below $370\,$nm is absorbed by the foil. Approximately $3-7.5\%$ of the light hitting the foil within the wavelength range $250\,\mathrm{nm} \leq \lambda \leq 390\,\mathrm{nm}$ is used for the WLS process. The intensity of the emission spectrum of the WLS light is slightly dependent on the absorbed wavelength and shows the shape of a rotational-vibrational fluorescence spectrum, peaking at around $\lambda \approx 420\,$nm. Adjusting the reflectance values to the measured ones in the Monte Carlo simulation originally used for the muon veto design, the veto detection efficiency remains unchanged. Including the wavelength shifting in the Monte Carlo simulation leads to an increase of the efficiency of approximately $0.5\%$.}


\begin{document}
\maketitle
\flushbottom

%\linenumbers

\section{Introduction} \label{sec:introduction}

The XENON1T experiment~\cite{Aprile_XENON1T_MC} aims for the direct detection of galactic dark matter in form of weakly interacting massive particles (WIMPs) or other types of dark matter, which arise from, e.g., supersymmetric theories. See for instance~\cite{Feng_DMCandidates}\cite{Aprile_XENON100_ALP} for a review. Based on the principle of a dual-phase time projection chamber (TPC) with liquid (LXe) and gaseous (GXe) xenon as detection medium, the detector is, like its predecessors XENON10~\cite{Aprile_XENON10_Inst} and XENON100~\cite{Aprile_XENON100_Inst}, operated at the Gran Sasso Underground Laboratory (LNGS) in Italy and is currently taking first science data. The goal of the experiment is to directly detect dark matter providing for this a projected sensitivity for the spin-independent WIMP-nucleon elastic cross-section of \mbox{$\approx2 \times 10^{-47}\,\mathrm{cm}^2$}~\cite{Aprile_XENON1T_MC}. Such an improvement of sensitivity requires, besides a $2\,\mathrm{t \cdot y}$ (ton-year) exposure, a background reduction of two orders of magnitude compared to XENON100. For a dark matter search experiment, the most dangerous kind of background are neutrons, due to the WIMP-like signature they produce when scattering off a nucleus. Neutrons can be produced through spontaneous fission (for instance from $^{238}$U), $\left(\alpha,n\right)$ reactions (radiogenic neutrons), or by cosmic muons interacting in the rock surrounding the detector (cosmogenic neutrons). To minimize the rate of radiogenic neutrons, highly purified water serves as passive shield against external radioactivities. Internal radioactivities are minimized by a careful selection of materials through a screening campaign prior to construction of the detector~\cite{Aprile_XENON1T}\cite{Lung_PMTs}\cite{Baudis_PMTs}. The cosmogenic neutron flux is first reduced by placing the experiment at the LNGS underground laboratory, where the muon flux is lowered by six orders of magnitude compared to the surface to a value of $\left(3.31 \pm 0.03\right)\cdot10^{-8}\,\mu/\mathrm{\left(cm^2s\right)}$~\cite{Selvi_MuonFlux} at an average muon energy of $270\,\mathrm{GeV}$~\cite{Ambrosio_MuonEnergy}. The detector itself is further surrounded by a 10 m high watertank filled with $750\,\mathrm{m^3}$ of continuously purified water and instrumented with $84$ $8\verb+"+$ photomultiplier tubes (PMTs) in order to work as an active Cherenkov muon veto.

The muon veto system has been optimized with a dedicated Monte Carlo (MC) study~\cite{Fattori_MV}: it yields a detection efficiency for muons crossing the water tank of about $99.5\%$ and $>70\%$ if just the secondary particles cross the tank. Furthermore, this MC study led to the choice of the DF2000MA foil by 3M as a reflector, covering the full inner surface of the water tank, in order to collect as much Cherenkov light as possible. The DF2000MA foil is an adhesive polymeric multilayer film with, according to the manufacturer, nearly $100\%$ reflectance for wavelengths larger than $430\,\mathrm{nm}$~\cite{3M}. A renewed data sheet states the high reflectance values at wavelengths larger $400\,\mathrm{nm}$ \cite{3M_2017}. 3M reflective foils have widely been used in rare event searches and other experiments, especially a foil named VM2000\footnote{As response to an inquiry, 3M classified the DF2000MA foil to be a successor of the VM2000 foil. We have no proprietary information on VM2000, or if it behaves the same as the DF2000MA foil inspected here (although it is expected).}, see e.g. \cite{GERDA} \cite{PANDA} \cite{WARP}. In addition, similar foils have shown (\cite{Angloher_CRESST}) wavelength shifting (WLS) capabilities, absorbing lower wavelength photons and re-emitting them with their wavelength shifted to higher values. This process could shift the UV part of the Cherenkov spectrum (which drops with $1/\lambda^2$) into the highly sensitive region of the PMTs, defined as where their quantum efficiency reaches values $> 10\%$ ($310\,\mathrm{nm}-540\,\mathrm{nm}$), hence increasing the detection efficiency of the muon veto.

In this work, we present a verification of the DF2000MA foil specular reflectance as well as a qualitative and quantitative inspection of its WLS properties. Reflectivity has been measured with dedicated setups at the Universities of Mainz and Erlangen. A fluorescence measurement at the University of Mainz led to absorption and emission spectra of the DF2000MA foil. The MC studies~\cite{Fattori_MV}, which were previously limited to the original reflectance values provided by 3M \cite{3M} and had not considered WLS, have been cross-checked with our measurement results including WLS. Finally, the effect of damages on the foil surface and the accompanied influence on the reflectance have been investigated.

\section{Foil Reflectivity} \label{sec:refl}

\subsection{Description of the Experimental Setup}\label{sec:refl:setup}

\begin{figure}[htb]
	\centering
	\includegraphics[width=0.70\textwidth]{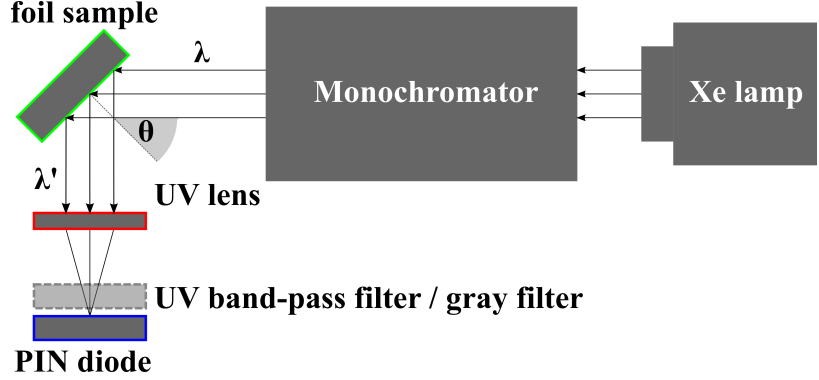}
	\caption{Setup to measure the specular reflectance of the DF2000MA foil. $\lambda$ expresses the wavelength selected at the monochromator. $\lambda'$ is the wavelength of light after reflection. For specular reflection one can assume $\lambda = \lambda'$.}
	\label{fig:refl:SetupSpecReflMainz}
\end{figure}

\begin{figure}[htb]
	\centering
	\subfigure[]{\label{fig:refl:PINdiode_Responsivity}\includegraphics[width=0.49\textwidth]{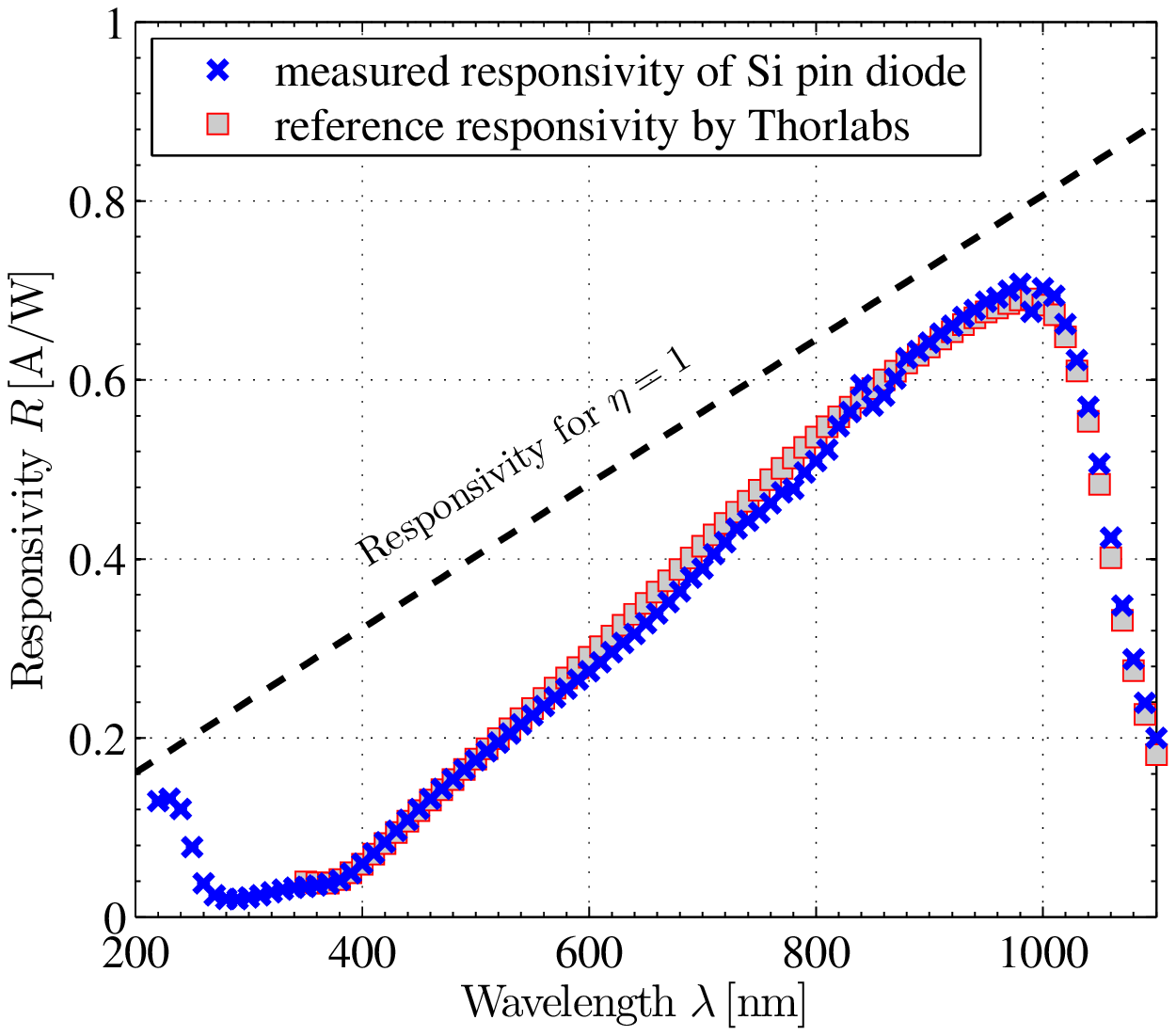}}	
	\subfigure[]{\label{fig:refl:SpectralGapwidth}\includegraphics[width=0.49\textwidth]{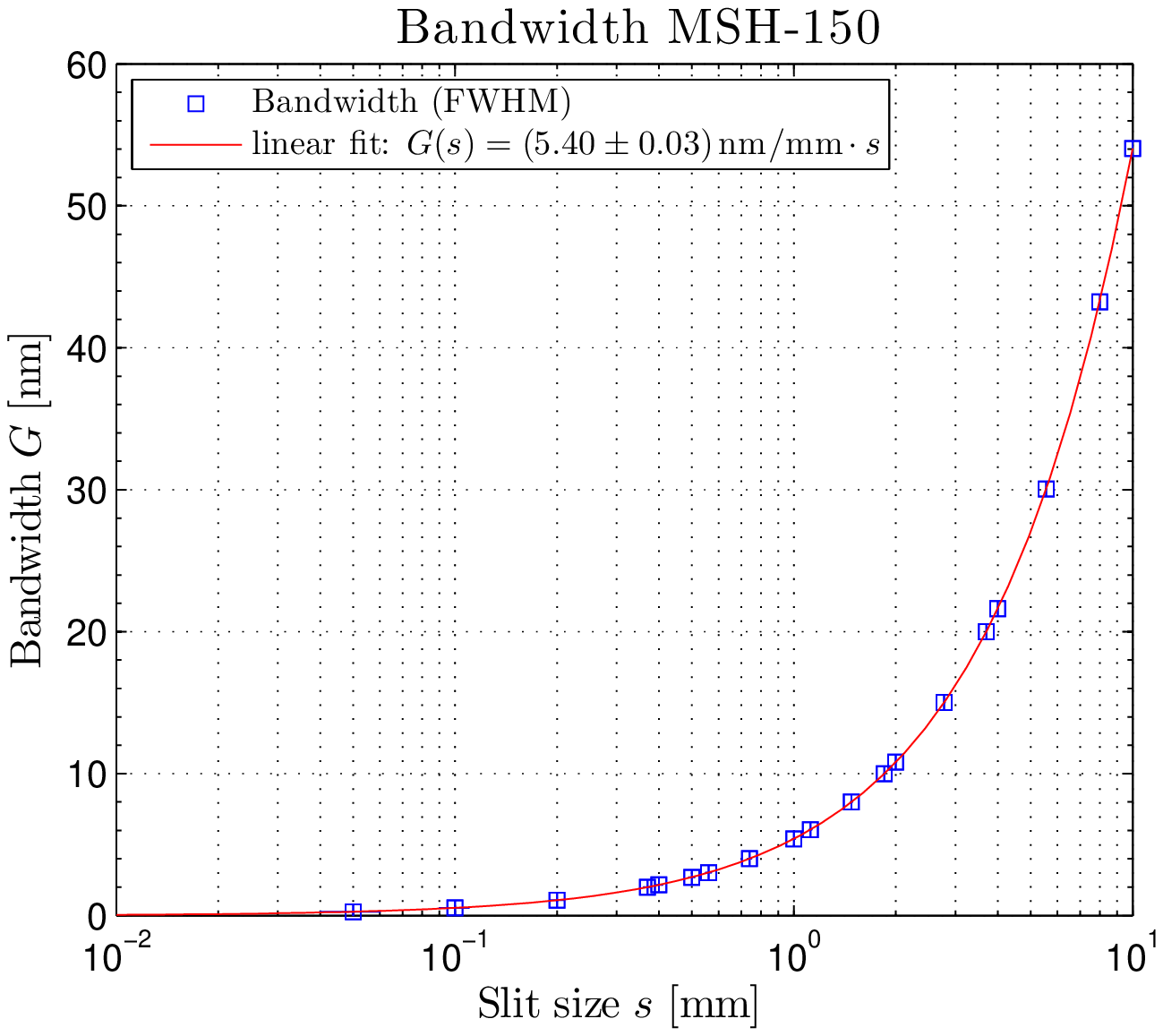}}
	\caption{{\em a)} The responsivity $R$ of the used Si PIN diode with values provided by ~\cite{Thorlabs_FDS1010-CAL} (red squares) and cross-check measurements done with a Ophir Vega power meter (\cite{Ophir}) (blue). The black line indicates the responsivity of an ideal PIN diode with a quantum efficiency $\eta = 1$; {\em b)} Bandwidth $G$ as a function of the selected slit size $s$ of the monochromator \cite{LOT}.}
  \label{fig:refl:SpectralGapwidth_and_PINdiode_Responsivity}
\end{figure}

The measurement takes place in a completely darkened room in which the optical setup is mounted on an optical table and is enclosed by a box with black-painted inner surfaces, to prevent possible residual light. A schematic illustration of the full setup is shown in figure~\ref{fig:refl:SetupSpecReflMainz}. A monochromator selects certain wavelengths $\lambda$ out of the spectrum of a continuous Xe arc lamp: the selected wavelength steps are $10\,\mathrm{nm}$, ranging from $280\,\mathrm{nm}$ to $580\,\mathrm{nm}$ covering all measurable wavelengths of Cherenkov light starting from the smallest wavelength transmitted by the used UV band-pass filter to the wavelength where the quantum efficiency of the muon veto PMTs drops below $5\%$. In the region between $370\,\mathrm{nm}$ and $400\,\mathrm{nm}$ the measurement was done in $5\,\mathrm{nm}$ steps, to increase precision in the region where the rise in the reflectance of the foil takes place. The light coming from the monochromator is sent to the sample holder (framed green in figure~\ref{fig:refl:SetupSpecReflMainz}), which is an aluminium cuboid covered with the DF2000MA foil on one side and a specular reflectance standard on the opposing side. This reflectance standard is the $1\verb+"+$ square PFSQ10 UV mirror by Thorlabs \cite{Thorlabs_PFSQ10}. The sample holder is rotatable and is attached to a goniometer that provides an angular precision of $0.5^{\circ}$, so that for each $\lambda$ the reflected light by the foil and by the standard can be measured under the same configuration. The reflected light is then collected via a lens (UV, framed red in figure~\ref{fig:refl:SetupSpecReflMainz}) and focussed on a calibrated Si PIN diode (framed blue on figure~\ref{fig:refl:SetupSpecReflMainz}) with known responsivity $R$ (FDS1010-CAL) \cite{Thorlabs_FDS1010-CAL} and a quadratic active area of $A = 100\,\mathrm{mm^2}$, which is placed in the beam path. The responsivity of a PIN diode is defined as the wavelength dependent ratio of its photocurrent $I_0$ to the power $P_0$ of the incident light: 

\begin{equation}
	R\left(\lambda\right) = \frac{I_0\left(\lambda\right)}{P_0\left(\lambda\right)} = \eta\left(\lambda\right) \frac{e}{hc} \lambda, \label{eqn:refl:Responsivity}
\end{equation}

with $e$ being the electron charge, $h$ Planck's constant and $c$ the speed of light. The quantum efficiency $\eta\left(\lambda\right)$ is the only wavelength dependent term. Figure~\ref{fig:refl:PINdiode_Responsivity} shows the responsivity of the diode over the wavelength. Inspected angles of incidence (AOI) were $\theta_1 = 45^\circ$ and $\theta_2 = 12^\circ$ for the whole measurement, matching the AOIs for which Thorlabs provides reflectance curves. We used an LA4148 UV lens by Thorlabs, which is uncoated and hence provides transmission values $> 95\%$ between $200\,\mathrm{nm}-1000\,\mathrm{nm}$ \cite{Thorlabs_LA4148}. During the measurements, it turned out that, depending on the wavelength selected, a UV band-pass filter or a gray filter in front of the PIN diode is useful (displayed as shaded box in figure~\ref{fig:refl:SetupSpecReflMainz}). The UV band-pass filter (FGUV11M \cite{Thorlabs_FGUV11M}) transmits wavelengths between $275\,\mathrm{nm}-375\,\mathrm{nm}$ and is necessary to discriminate against light which has been wavelength shifted by the foil to longer wavelengths (see section~\ref{sec:WLS}). The gray filter (NE10B \cite{Thorlabs_NE10B}) reduces the intensity of the incoming light by transmitting only $\left(10 \pm 2\right)\%$ within the wavelength range it was used in: $410\,\mathrm{nm}-580\,\mathrm{nm}$. This is necessary since the current values measured by the PIN diode become very high for wavelengths in that part of the spectrum. The diode current is measured for a certain angle of incidence, once for the DF2000MA foil and once for the specular reflectance standard: the comparison of these two measured current values provides an estimate of the reflectance curve of the DF2000MA foil (see section ~\ref{sec:refl:analysis}).

The used monochromator is the MSH-150 by LOT \cite{LOT} in Czerny-Turner setup, designed for the ultraviolet and optical bands of the electromagnetic spectrum. Both, Xe lamp and monochromator are outside of the dark box. The incoming light is dispersed by focusing the Xe light, behind the input slit, with a hollow mirror on a table-mounted diffraction grating. The diffracted light is reflected by another hollow mirror onto the output slit. Rotation of the grating table leads to a different dispersion and to the selection of the desired wavelength window. The bandwidth $G$ of the monochromator is defined as the wavelength representing full width at half maximum (FWHM) of the spectral output light intensity profile, which has according to \cite{LOT} a triangular shape for small slit sizes ($\ll 0.5\,\mathrm{mm}$) and a rectangular shape for large slit sizes ($\gg 0.5\,\mathrm{mm}$). The slit sizes are adjustable by a micrometer screw gauge. For this measurement, the slit size for both slits was selected to be $0.07\,\mathrm{mm}$, which translates to $G = 0.38\,\mathrm{nm}$ according to \cite{LOT}. The output light of the monochromator is thus described by a rectangle function centered at the selected wavelength $\lambda$ with a width of $G$. Figure~\ref{fig:refl:SpectralGapwidth} shows that the relation between $G$ and the slit size $s$ can be fit well ($\chi^2_\mathrm{red} = 1.27$) by a linear function including systematic uncertainties of $\Delta s = 10\,\mathrm{\upmu m}$ (precision of the micrometer screw gauge) and $\Delta G = 0.01\,\mathrm{nm}$ (accuracy of the bandwidth according to \cite{LOT}). Systematics induced by this bandwidth to the measurement presented here, were taken into account (see section~\ref{sec:refl:analysis}). 

\begin{figure}[htb]
	\centering
	\includegraphics[width=0.70\textwidth]{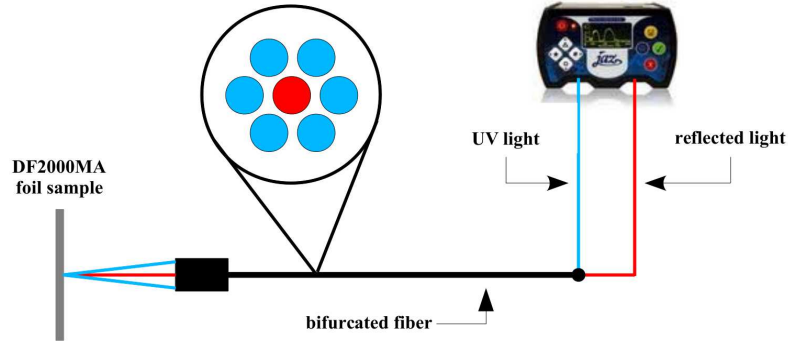}
	\caption{Verification setup including the Ocean Optics JAZ spectrometer. Light is guided to the probe surface (blue), where it gets reflected and guided back to the module (red).}
	\label{fig:refl:SetupSpecReflErlangen}
\end{figure}

To verify the results obtained by the setup described above, a second measurement setup was arranged including the JAZ spectrometer by Ocean Optics \cite{OceanOptics}: this device is developed for spectrographic analysis of surfaces. The used model is composed of a display unit, an interface module to communicate with computers via ethernet, a light source module containing a pulsed Xe arc lamp and the lattice based spectrometer module which is able to distribute incoming light onto $2048$ CCD pixels. The number of hits on a CCD pixel is a measure of the light intensity for the corresponding wavelength range. The selectable wavelength range of the device is $250 \leq \lambda \leq 800\,\mathrm{nm}$. The sketch in figure~\ref{fig:refl:SetupSpecReflErlangen} shows how the light is guided from the module to the probe surface through a bifurcated fiber consisting of seven single quartz fibers (diameter of $600\,\mathrm{\upmu m}$) arranged such that six fibers symmetrically surround a central one. The light produced in the JAZ module is 
guided to the probe through the outer six fibers, gets reflected on the surface and collected by the central fiber at an angle of incidence of $0^\circ$, which guides the light back to the spectrometer.

The spectrometer module is connected to a computer and controlled by a dedicated software. The JAZ spectrometer was calibrated using the reflectance standard UV mirror described previously. The software compares the reflectance of the probe surface with the known reflectance of the UV mirror and calculates the specular reflectance value for a given wavelength. Several DF2000MA foil samples (area of $\approx15\,\mathrm{cm^2}$) were made and inspected with the JAZ spectrometer:

\begin{itemize}
 \item an undamaged piece of DF2000MA foil to measure the reflectance curve as well as to cross-check the monochromator measurement with the clean sample;
 \item a piece with half-width of foil slightly scratched and another half-width of foil heavily scratched to estimate the impact of a foil surface damage on its reflectance;
\end{itemize}

\subsection{Data and Analysis}\label{sec:refl:analysis}

The reflective properties of the foil for a thick sample is expressed by the power ratio between the reflected light and the light before reflection

\begin{equation}
	\rho = \frac{P_{refl}}{P_0}. \label{eqn:refl:Reflectance}
\end{equation}

The value $\rho$ is called {\em reflectance} and is usually expressed in units of percent. To derive it, the PIN diode current has to be measured for each reflected wavelength. To assess the optical input power $P_0$, the light source spectrum $S\left(\lambda\right)$ can be measured in $\mathrm{\upmu A}$ at the output of the monochromator with the PIN diode. According to the setup described in section~\ref{sec:refl:setup}, the wavelength dependent photocurrent $I\left(\lambda\right)$ of the PIN diode can be described as a convolution of the light source spectrum $S\left(\lambda\right)$, measured with the PIN diode, and the wavelength dependent reflectance transfer function $M\left(\lambda\right)$ of the DF2000MA foil or the UV mirror, scaled with the responsivity $R\left(\lambda'\right)$ of the PIN diode, where $\lambda$ describes the selected wavelength at the monochromator and $\lambda'$ describes the wavelength of the light after being reflected by foil or mirror. The systematics of the experimental setup are expressed with the variable $\Lambda\left(G\right) = C\cdot G$, where $C$ is a setup specific constant with the dimension $\left[C\right] = \mathrm{W/\left(A\cdot nm\right)}$ and $G\left(s\right)$ is the monochromators bandwidth (see figure~\ref{fig:refl:SpectralGapwidth}). $\Lambda$ has the dimension $\left[\Lambda\right] = \mathrm{W/A}$, such that $I\left(\lambda\right) \propto \Lambda\left(G\right)$: 

\begin{equation}
  \begin{split}
    I\left(\lambda\right) &= \Lambda\left(G\right) \cdot R\left(\lambda'\right) \cdot L\left(\lambda'\right) \cdot \left(M\left(\lambda\right) \ast S\left(\lambda\right)\right) \\     
	       &= \Lambda\left(G\right) \cdot R\left(\lambda\right) \cdot L\left(\lambda\right) \int_{\lambda} M\left(\widetilde{\lambda}\right) \cdot S\left(\lambda-\widetilde{\lambda}\right) \, \mathrm{d} \widetilde{\lambda},            
  \end{split}
  \label{eqn:refl:Convolution_SpecularReflection_calculation}
\end{equation}

with $L\left(\lambda\right)$ being the transfer function of the lens. Since specular reflection does not change the wavelengths before and after the foil ($\lambda' = \lambda$), the responsivity remains the same: $R\left(\lambda'\right) = R\left(\lambda\right)$. Further, $L\left(\lambda'\right) = L\left(\lambda\right)$ and $M\left(\lambda\right)$ can be simply expressed as $\rho\left(\lambda\right)$. The very small bandwidth $G = 0.38\,\mathrm{nm}$ allows to express the light from the source as $S\left(\lambda-\widetilde{\lambda}\right) = S_0\left(\lambda\right) \cdot \delta\left(\lambda-\widetilde{\lambda}\right)$, where $\delta\left(\lambda\right)$ is Diracs delta function at $\lambda$. Applying the identiy of the Dirac function, makes the convolution collapse to a simple product

\begin{equation}
  \begin{split}
    \displaystyle I\left(\lambda\right) &= \Lambda\left(G\right) \cdot R\left(\lambda\right) \cdot L\left(\lambda\right) \int_{\lambda} \rho\left(\widetilde{\lambda}\right) \cdot S_0\left(\lambda\right) \cdot \delta\left(\lambda-\widetilde{\lambda}\right) \, \mathrm{d}\widetilde{\lambda} \\
	       &= \Lambda\left(G\right) \cdot R\left(\lambda\right) \cdot L\left(\lambda\right) \cdot \rho\left(\lambda\right) \cdot S_0\left(\lambda\right).
  \end{split}
  \label{eqn:refl:Convolution_SpecularReflection}
\end{equation}

\begin{figure}[htb]
	\centering
	\includegraphics[width=0.70\textwidth]{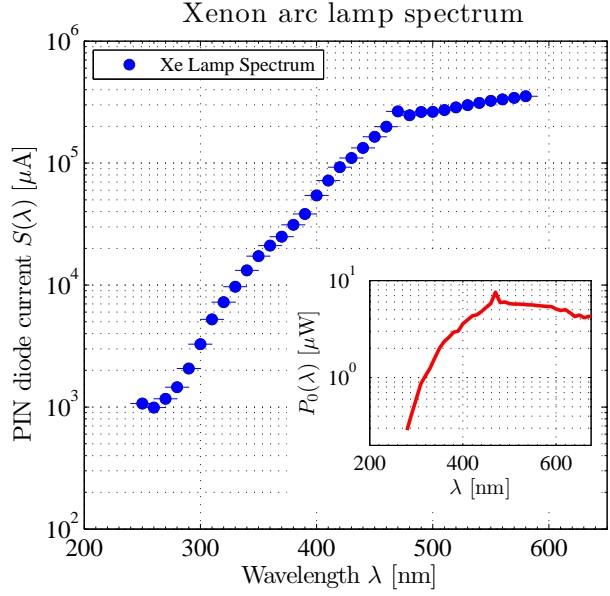}
	\caption{The spectrum $S\left(\lambda\right)$ of the xenon arc lamp measured by the PIN diode (blue dots) in $\upmu$A and measured by a calibrated power meter (red, shown in the lower right corner) in $\upmu$W. A xenon spectral line at $484.4\,\mathrm{nm}$ can be seen.}
	\label{fig:refl:XeLampSpectrum}
\end{figure}

The spectrum of the light source $S\left(\lambda\right)$ was measured with the PIN diode in units of $\mathrm{\upmu A}$ and is shown in figure~\ref{fig:refl:XeLampSpectrum} together with a measurement $P_0$ of the spectrum obtained by the same power meter used for figure~\ref{fig:refl:PINdiode_Responsivity}. Both are according to \eqref{eqn:refl:Responsivity} connected by

\begin{equation}
	R = \frac{S\left(\lambda\right)\left[\upmu\mathrm{A}\right]}{P_0\left(\lambda\right)\left[\upmu\mathrm{W}\right]}. \label{eqn:refl:Responsivity_spectrum}
\end{equation}

\begin{figure}[htb]
	\centering
	\includegraphics[width=0.70\textwidth]{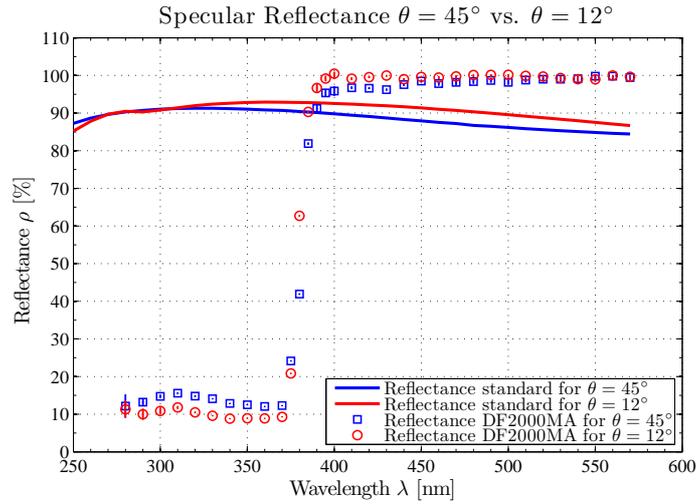}
	\caption{Specular reflectance $\rho$ of the DF2000MA foil for the two inspected angles of incidence ($\theta_1 = 45^\circ$ blue boxes, $\theta_2 = 12^\circ$ red circles) versus the wavelength $\lambda$. The thicker blue and red lines show the reflectance curve of the used specular reflectance standard at $\theta_1$ and $\theta_2$, respectively.}
	\label{fig:refl:AnalysisResultsMainz}
\end{figure}

With help of \eqref{eqn:refl:Convolution_SpecularReflection} it is possible to derive $\rho\left(\lambda\right)$ of the foil using the measured PIN diode currents $I\left(\lambda\right)$ and $I_m\left(\lambda\right)$ for foil and mirror, respectively, as well as the known reflectance curve of the UV mirror $\rho_m$. For angles of incidence equal to $12^\circ$ and $45^\circ$, the reflectance of the DF2000MA foil can be thus calculated by:

\begin{equation}
	\rho\left(\lambda\right) = \frac{I\left(\lambda\right)}{I_{m}\left(\lambda\right)} \cdot \rho_m\left(\lambda\right). \label{eqn:refl:SpecularReflectance}
\end{equation}

which is independent of source spectrum $S\left(\lambda\right)$, the responsivity of the PIN diode $R\left(\lambda\right)$ as well as the lens transfer function $L\left(\lambda\right)$. The result of this analysis is shown in figure~\ref{fig:refl:AnalysisResultsMainz}. The edge of the curve, where the reflectance rises, is already observed at $\lambda = 370\,\mathrm{nm}$ and not at $400\,\mathrm{nm}$ as specified by 3M~\cite{3M}. The rise of reflectance ends at $400\,\mathrm{nm}$. There is a slight dependence of the reflectance on the angle of incidence: at $12^\circ$, the reflectance is $>99\%$ above $400\,\mathrm{nm}$, whereas at $45^\circ$, the reflectance rises from $96\%$ at $400\,\mathrm{nm}$ to $99\%$ at $450\,\mathrm{nm}$. For wavelengths $<370\,\mathrm{nm}$ the foil reflects about $13\%$ of the incoming light at $45^\circ$ and about $10\%$ at $12^\circ$. The error on the reflectance values is at $\approx0.5\%$ increasing up to $\approx 2\%$ for $\lambda < 300\,\mathrm{nm}$.

\begin{figure}[htb]
  \centering
  \includegraphics[width=0.70\textwidth]{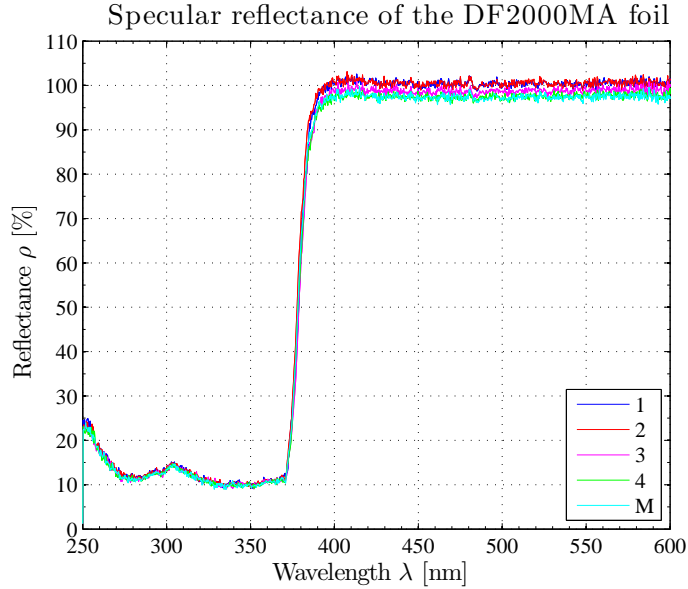}
  \caption[]{The specular reflectance of the DF2000MA foil as measured with the JAZ spectrometer at $0^\circ$ incidence angle near the four corners and at the center (M) of the sample. It is nearly uniform over the foil surface (the measurements on five positions lead to a difference of $2\%-3\%$). }
  \label{fig:refl:AnalysisResultsErlangenDifferentPointsCurves}
\end{figure}

The cross check measurements with the JAZ spectrometer module were performed at five different spots on the foil probe in order to verify if equal reflectance can be ensured over the whole surface (positions $1$ to $4$ in the corners of the sample and position $M$ in the middle). The results of these measurements are plotted in figure~\ref{fig:refl:AnalysisResultsErlangenDifferentPointsCurves}, in which a clear consistency is visible: the maximum difference of the curves for $\lambda > 400\,\mathrm{nm}$ is about $2\%-3\%$ which is, in case of the XENON1T Muon Veto, negligible~\cite{Fattori_MV}.

In addition, the measurements confirm the results obtained with the monochromator setup (figure~\ref{fig:refl:AnalysisResultsMainz}): the rise of reflectance begins at $370\,\mathrm{nm}$, is $30\,\mathrm{nm}$ wide and reaches nearly $100\%$ at $400\,\mathrm{nm}$. Down to $275\,\mathrm{nm}$ this curve at $0^\circ$ incidence angle looks similar to the curve measured in figure~\ref{fig:refl:AnalysisResultsMainz} at $12^\circ$ incidence.

\begin{figure}[htb]
  \centering
  \includegraphics[width=0.70\textwidth]{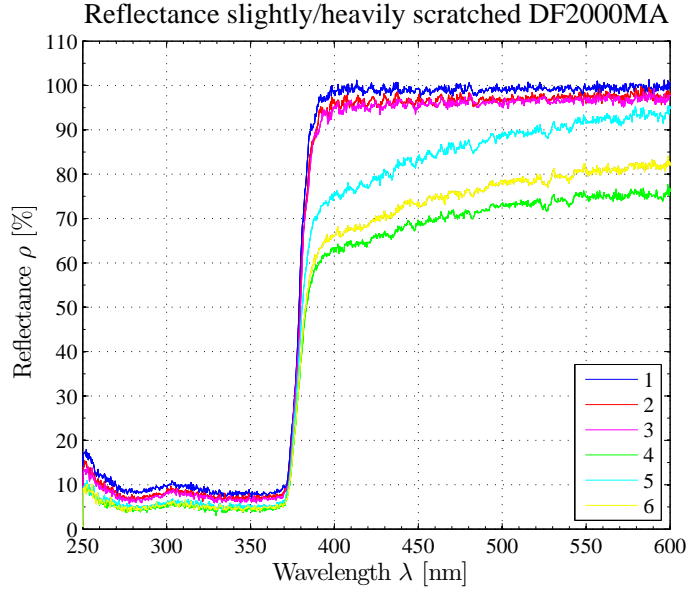}
  \caption[]{Reflectance of a DF2000MA foil sample with intentional scratches. While slight scratches (positions $1$ to $3$) just cause minor decrease of the reflectance, a heavily scratched surface (positions $4$ to $6$) leads to a drop of reflectance down to $\approx60\%$ compared to the original value with decreasing impact at longer wavelengths.}
  \label{fig:refl:AnalysisResultsErlangenScratchesCurves}
\end{figure}

To assess the potential impact of surface damages on the foil reflectance qualitatively, an altered foil sample with damages as described in section~\ref{sec:refl:setup} was inspected. On each half, measurements at three different points have been performed. The results in figure~\ref{fig:refl:AnalysisResultsErlangenScratchesCurves} show, that the slightly damaged half sample still provides comparable reflectance to the values shown in figure~\ref{fig:refl:AnalysisResultsErlangenDifferentPointsCurves} for a non-damaged sample. A small drop of about $5\%$ over the whole measurement wavelength interval can be observed for the red and magenta curves (positions 2 and 3). However, for the heavily damaged half of the sample, a drop is clearly visible. Depending on the position of the measurement and the reflected wavelength, the reflectance just reaches $60\%-90\%$ of the value for a non-damaged foil sample. It seems that deep scratches destroy the reflecting polymer layer, causing a large drop in reflectance. Thus, during construction of the XENON1T muon veto, the cladding of the DF2000MA foil was performed with special care and no deep scratches were inflicted to the foils surface.

\section{Foil Wavelength Shifting Properties} \label{sec:WLS}

The DF2000MA foil has also wavelength shifting (WLS) properties. Photons hitting the foil can get absorbed and reemitted with a higher wavelength through the process of fluorescence. When a molecule of one of the polymeric layers of the foil gets excited by the absorbed light, one part of the excitation energy is transformed into heat while another part is emitted as photons of a longer wavelength in diffuse directions. This can be a useful property to shift ultraviolet Cherenkov light to a higher wavelength regime, where the quantum efficiency of the XENON1T muon veto PMTs is higher \cite{Fattori_MV}.

\begin{figure}[htb]
  \centering
  \includegraphics[width=0.70\textwidth]{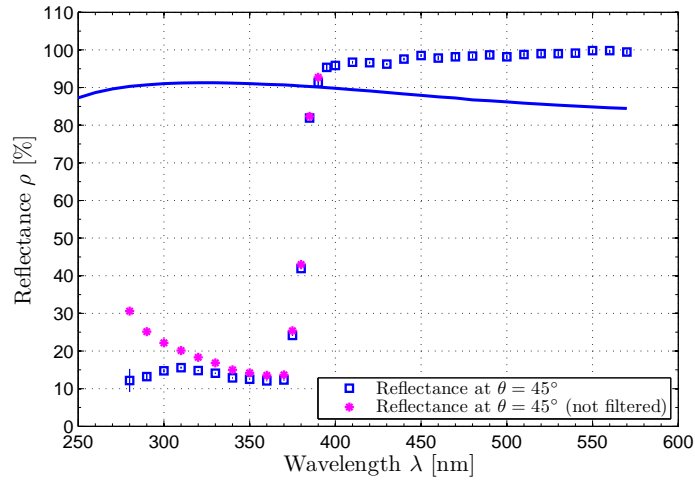}
  \caption[]{First indication of wavelength shifting properties of the DF2000MA foil. Measured values below $370\,\mathrm{nm}$ are higher without the band-pass filter (purple dots). Blocking the wavelength shifted light with the UV band-pass filter leads to the real specular reflectance curve of the foil.}
  \label{fig:WLS:WLS_Motivation}
\end{figure}

A first proof that the foil has indeed wavelength shifting properties can be obtained from the specular reflectance measurement. Figure~\ref{fig:WLS:WLS_Motivation} shows the same curve as it is plotted in figure~\ref{fig:refl:AnalysisResultsMainz} for two different conditions. Once with the UV band-pass filter (blue squares), described in section~\ref{sec:refl:setup}, and once without (purple dots). It is visible that without the UV band-pass filter the PIN diode current is higher for wavelengths below $370\,\mathrm{nm}$. This indicates that light from this wavelength interval got wavelength shifted to larger wavelengths, resulting in a higher PIN diode current. 

\subsection{Description of the Experimental Setup}\label{sec:WLS:setup}

% \subsubsection{pin Diode Setup}\label{sec:WLS:pinsetup}

\begin{figure}[htb]
	\centering
	\includegraphics[width=0.70\textwidth]{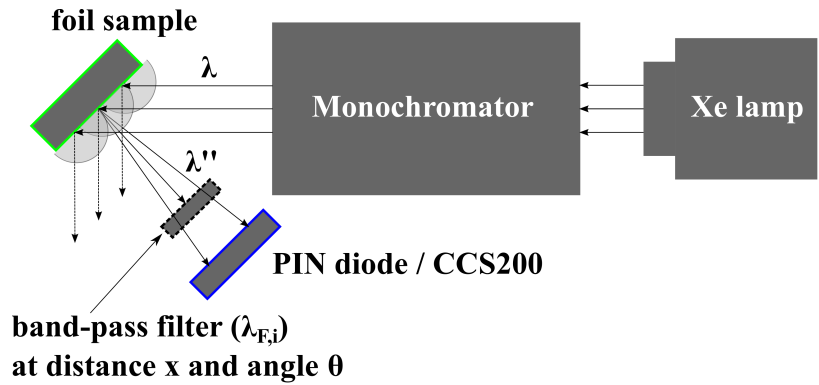}
	\caption{Setup to measure the wavelength shifting properties of the DF2000MA foil. $\lambda$ stands again for the wavelength selected at the monochromator. Additional band-pass filters transmitting certain wavelengths $\lambda_{F,i}$ are located at disctance $x$ and angle $\theta$ to quantify the diffusively emitted WLS light $\left(\lambda''\right)$ for the PIN diode measurement. For the measurement with the CCS200 spectrometer no band-pass filters are used.}
	\label{fig:WLS:SetupWLSMainz}
\end{figure}

To characterise the wavelength shifting property of the foil, the setup described in section~\ref{sec:refl:setup} was modified: the PIN diode was shifted out of the specular beam path to make the measurement independent of the specular reflectance of the foil. Due to the low intensity of the light outside of the beam path, the output slit of the monochromator was opened to $2\,\mathrm{mm}$  which corresponds, according to figure~\ref{fig:refl:SpectralGapwidth}, to a bandwidth of $10.81\,\mathrm{nm}$. The inspected wavelength interval was $250\,\mathrm{nm} \leq \lambda \leq 580\,\mathrm{nm}$. The modified setup is shown schematically in figure~\ref{fig:WLS:SetupWLSMainz}. 

The distance between foil surface and filter was chosen to be $x = \left(2.0 \pm 0.1\right)\,\mathrm{cm}$, the PIN diode is located directly, $x' = \left(2 \pm 1\right)\,\mathrm{mm}$, behind. Thus, in this setup, the diode sees a certain solid angle $\Omega = A/\left(x+x'\right)^2$ of the diffusively emitted light, with $A$ being the PIN diodes effective surface. The measured current values have been corrected for the full solid angle during analysis. Similar to the specular reflectance measurement, the current values of the diode were measured once for the DF2000MA foil and once for a diffuse reflectance standard. The reflection standard is a $2.5\,\mathrm{cm}$ diameter disk of porose polytetrafluoroethylene (PTFE)  with a lambertian behaviour, a known reflectance curve $\rho_{mf}\left(\lambda\right)$ and without wavelength shifting properties \cite{SphereOptics}. 

To quantify the diffuse reflectance and the WLS, five optical band-pass filters (Edmund Optics NT84 series \cite{EdmundOptics} and the FGUV11M filter mentioned above) are used in front of the PIN diode. They transmit only wavelengths $\lambda_{F,i}$ within their bandwidth (see table~\ref{tab:WLS:Filters}). By changing these filters, it is possible to measure the amount of light shifted to higher wavelengths into the corresponding five possible bandwidths. By inspecting those five intervals for each wavelength $\lambda$ selected at the monochromator, it is possible to cover the whole range in which the muon veto PMTs have their highest quantum efficiency. If not specified explicitly, all following plots and results originate from one measurement run of all wavelengths $\lambda$ measured five times seperately over each filter. Since the surface of the DF2000MA foil has high specular reflection (see section~\ref{sec:refl}, the expected diffuse reflectance is quite low. Nevertheless, if the selected wavelength $\lambda$ is within one of the bandwidths of the filters, one can quantify the diffuse reflectance of the foil, by integrating all values inside that single filter bandwidth. 

While the setup with the PIN diode is expected to give a good quantitative result of the WLS, it is only giving limited spectroscopic results of the WLS spectrum since its resolution of the emission wavelength is limited to the bandwidth of the used filters ($50\,\mathrm{nm}$ and $100\,\mathrm{nm}$, respectively). To measure the spectrum of the WLS with higher resolution, the Thorlabs CCS200 Spectrometer~\cite{Thorlabs_CCS200} was used by replacing the PIN diode in the setup (as indicated in figure~\ref{fig:WLS:SetupWLSMainz}) and the band-pass filters were removed. For the measurement, the spectrometer was placed such that its light input (open SMA connector) was placed $2\,$cm away from the foil sample, perpendicular to the surface of the sample. The possibility of an optical fiber connected to the spectrometers input was not used in order to maximize the amount of light going into the spectrometer. The CCS200 is a compact lattice spectrometer with a spectroscopic resolution of $\approx0.25\,\mathrm{nm}$ and a nominal measurement range of $200\,\mathrm{nm}-1000\,\mathrm{nm}$. However, its factory calibration is only done with a light source which does not have a high intensity in the deep UV. Therefore, its calibration values are only reliable for wavelengths greater than $375\,\mathrm{nm}$. Moreover, the calibrated intensity values measured are normalized to the whole intensity hitting the spectrometer, which is not available. So one can obtain a very good spectroscopic measurement, but a limited absolute intensity measurement. Due to these limitations, a dedicated analysis has to be performed to compare the measurement values obtained with the PIN diode quantitatively with those obtained with the CCS200 spectrometer.

\subsection{Data and Analysis}\label{sec:WLS:analysis}

Each one of the optical band-pass filters has a transmittance $\tau$ depending on the central wavelength $\lambda_{c,i}$ and a bandwidth of $50\,$nm, respectively $100\,\mathrm{nm}$, making it transparent to wavelengths $\lambda_{F,i}$. The values can be extracted from their data sheets \cite{Thorlabs_FGUV11M},\cite{EdmundOptics} and are listed in table~\ref{tab:WLS:Filters}.

\begin{table}[htb!]
  \centering
  \begin{tabular}{@{}cccccc@{}} \toprule
    \multicolumn{6}{c}{Band-pass Filters} \\ \cmidrule(r){1-6}
    \textbf{Filter} 	& \textbf{Bandwidth}		& ${\bm \tau}$\,\textbf{[\%]}		& ${\bm \lambda_{c,i}}$\,\textbf{[nm]}	& \textbf{FWHM\,[nm]}	& ${\bm R\left(\lambda_{F,i}\right)}$\,\textbf{[A/W]}\\ 
    $1$     		& $275\,\mathrm{nm} \leq \lambda_{F,1} \leq 375\,$nm	& $80 \pm 5$	& $325$						& $100$			& $0.028 \pm 0.007$			\\
    $2$     		& $375\,\mathrm{nm} \leq \lambda_{F,2} \leq 425\,$nm	& $90 \pm 5$	& $400$						& $50$			& $0.061 \pm 0.017$			\\ 
    $3$     		& $425\,\mathrm{nm} \leq \lambda_{F,3} \leq 475\,$nm	& $95 \pm 2$	& $450$						& $50$			& $0.120 \pm 0.019$			\\ 
    $4$     		& $475\,\mathrm{nm} \leq \lambda_{F,4} \leq 525\,$nm	& $95 \pm 2$	& $500$						& $50$			& $0.175 \pm 0.016$			\\ 
    $5$     		& $525\,\mathrm{nm} \leq \lambda_{F,5} \leq 575\,$nm	& $95 \pm 2$	& $550$						& $50$			& $0.225 \pm 0.016$			\\ 
    \bottomrule
  \end{tabular}
  \caption{The used band-pass filters and their corresponding bandwidths, transmittances $\tau$ within the bandwidth, central wavelengths $\lambda_{c,i}$, FWHMs and the mean PIN diode responsivities $R\left(\lambda_{F,i}\right)$ within the bandwidths.}
  \label{tab:WLS:Filters}  
\end{table}

During the analysis of the data taken in the WLS measurement, these filter systematics have been taken into account. In particular, the PIN diode current $I\left(\lambda\right)$ is now also dependent on the transmitted filter wavelengths $\lambda_{F,i}$: a convolution of the current with the transmission function $F\left(\lambda_{F,i}\right)$ of the filters needs to be added when expressing its wavelength dependence. Hence, the modified equation~\eqref{eqn:refl:Convolution_SpecularReflection_calculation} for each filter $i$ can be written as:

\begin{equation}
  \begin{split}
    I\left(\lambda,\lambda_{F,i}\right) &= \Lambda\left(G\right) \cdot R\left(\lambda''\right) \cdot \left\{F\left(\lambda_{F,i}\right) \ast \left[M\left(\lambda,\lambda_{F,i}\right) \ast S\left(\lambda\right)\right]\right\}. \\
  \end{split}
  \label{eqn:WLS:Convolution_WLS_calculation}
\end{equation}

$S\left(\lambda\right)$ and $\Lambda\left(G\right)$ express the spectrum and systematics of the light source, respectively. $\lambda$ is the chosen wavelength at the monochromator. The optical response function of the DF2000MA foil $M\left(\lambda,\lambda_{F,i}\right)$ is now dependent on two different wavelengths, because the wavelengths $\lambda''$ of the WLS light emitted within wavelengths $\lambda_{F,i}$ have now to be taken into account. The edges of the filter functions are sharp enough to be described as a rectangle with height $\tau_i$ and their FWHM as width (see filter curves provided by \cite{EdmundOptics}), such that the convolution with $F\left(\lambda_{F,i}\right)$ can be simplified to a multiplication with $\tau_i$.

\begin{equation}
  \begin{split}
    I\left(\lambda,\lambda_{F,i}\right) &= \Lambda\left(G\right) \cdot R\left(\lambda_{F,i}\right) \cdot \tau_i \cdot \left(M\left(\lambda,\lambda_{F,i}\right) \ast S\left(\lambda\right)\right) \\
  \end{split}
  \label{eqn:WLS:Convolution_WLS_calculations_1}
\end{equation}

Since for this setup the bandwidth $G = 10.81\,$nm, $S\left(\lambda\right)$ can not be described as a delta function anymore, but has to be described using a rectangle function $\Pi\left(\lambda\right)$ centered at $\lambda_0$ with the width $G$:

\begin{equation}
  \begin{split}
    I\left(\lambda,\lambda_{F,i}\right) &= \Lambda\left(G\right) \cdot R\left(\lambda_{F,i}\right) \cdot \tau_i \cdot \left(M\left(\lambda,\lambda_{F,i}\right) \ast S_0\left(\lambda\right) \cdot \Pi\left(\frac{\lambda-\lambda_0}{G}\right)\right) \\
			     &= \Lambda\left(G\right) \cdot R\left(\lambda_{F,i}\right) \cdot \tau_i \cdot \left\{M\left(\lambda,\lambda_{F,i}\right) \ast S_0\left(\lambda\right) \cdot \left[\Theta\left(\lambda - \left(\lambda_0-\frac{G}{2}\right)\right) - \Theta\left(\lambda - \left(\lambda_0+\frac{G}{2}\right)\right)\right]\right\},
  \end{split}
  \label{eqn:WLS:Convolution_WLS}
\end{equation}

where $\Theta\left(\lambda\right)$ is Heaviside's step function. This equation is analytically not further simplifiable. However, if one assumes that the directivity of the WLS light is equally probable in every direction, as for the diffuse reflectance standard, it is possible to compare the current measurements for each filter directly to each other. In that case the diffuse response $M\left(\lambda,\lambda_{F,i}\right)$ can be expressed as $\rho_{f,i}\left(\lambda,\lambda_{F,i}\right)$ which depends on the absorption wavelength $\lambda$ and the emission wavelength $\lambda''$ falling into the range of transmitted wavelengths $\lambda_{F,i}$ of the used filter to account for the wavelength shitfted light. For the diffuse reflectance standard, where $\lambda = \lambda''$ (no WLS), equations \eqref{eqn:WLS:Convolution_WLS_calculation} and \eqref{eqn:WLS:Convolution_WLS} result in a $\rho_{mf}$ only depending on the absorption wavelength. A measurement without filters, however, measures the PIN diode current $I\left(\lambda\right)$ over all wavelengths. Thus, the global diffuse response $\rho_f$ of the DF2000MA foil is

\begin{equation}
	\rho_f\left(\lambda\right) = \frac{ I\left(\lambda\right)}{I_{mf}\left(\lambda\right)} \cdot \rho_{mf}\left(\lambda\right), \label{eqn:WLS:DiffuseResponse}
\end{equation}

where $\rho_f$ is a combination of diffuse reflectance $\rho_{f,0}$ and the percentage of how much light gets wavelengthshifted (the latter one is called WLS ratio in the following).  

\begin{figure}[htb]
	\centering
	\includegraphics[width=0.70\textwidth]{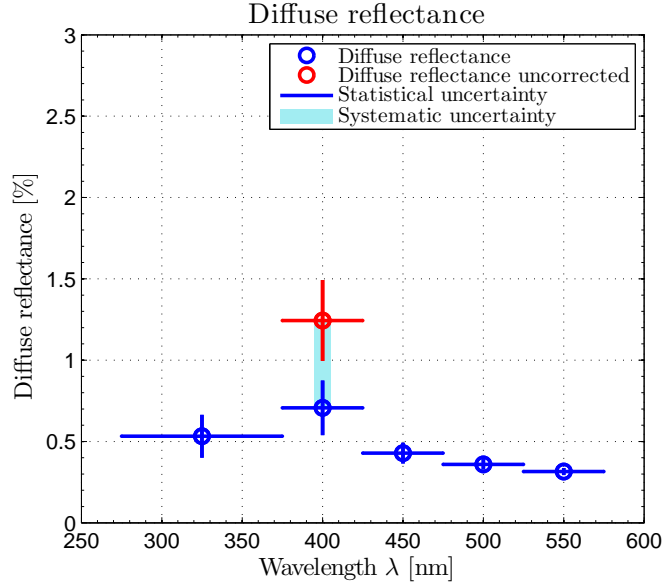}
	\caption{The diffuse reflectance $\rho_{f,0}$ of the DF2000MA foil. Values up to $0.7\%$ can be reached when integrating individually over the full filter bandwidths (blue circles). A correction for the contribution of WLS at an absorption wavelength of $380\,\mathrm{nm}$ has been performed. The corresponding value includes a systematic uncertainty as indicated by the blue shaded region. If not corrected for, a higher value for the second data point is obtained (red circle).}
	\label{fig:WLS:REALDiffuseReflectanceCurve}
\end{figure}

A summation of all measurement values at which the monochromator wavelength $\lambda$ is within the transmission window of each single filter leads to five datapoints. From those one can determine the diffuse reflectance $\rho_{f,0}$ of the foil with a resolution given by the bandwidth of the filters. Other light, eventually wavelength shifted to a $\lambda''$ outside of the corresponding filters transmission window, is blocked and has no influence on the PIN diode current. The result is shown in figure~\ref{fig:WLS:REALDiffuseReflectanceCurve}. One can see that there is a small component of diffuse reflection in the order of $1$\% depending on the wavelength interval. The value for the second data point is too high compared to the others. This is due to a systematic effect induced by the measurement at an absorption wavelength of $\lambda = 380\,\mathrm{nm}$, which has to be corrected for. This systematics are assessed in section~\ref{sec:systematics:WLS}.

\begin{figure}[htb]
 \centering
 \subfigure[]{\label{fig:WLS:Emissionsspectrum_300nm}\includegraphics[width=0.426\textwidth]{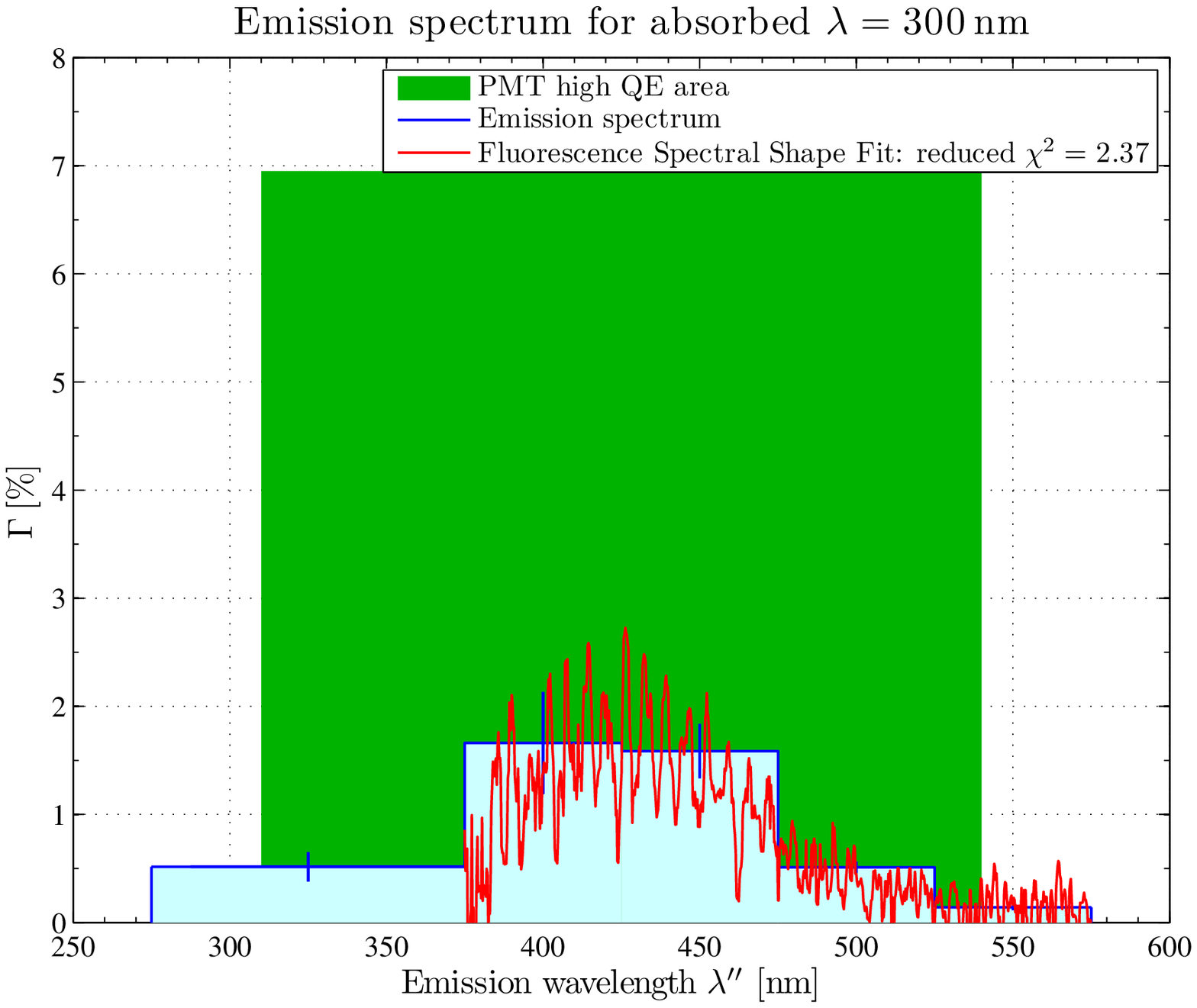}} 
 \hfill
 \subfigure[]{\label{fig:WLS:Emissionsspectrum_320nm}\includegraphics[width=0.426\textwidth]{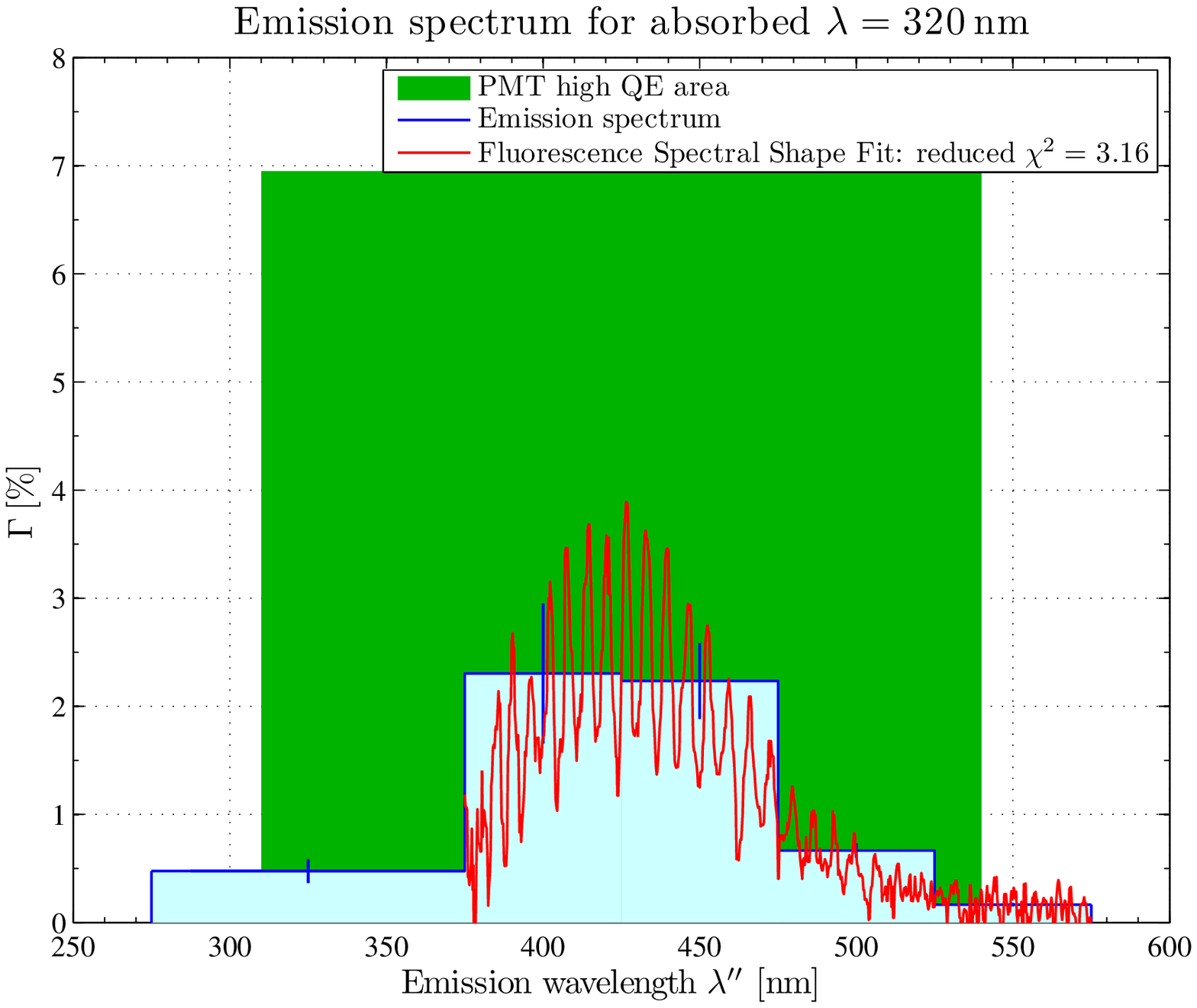}} 
 \hfill
 \subfigure[]{\label{fig:WLS:Emissionsspectrum_370nm}\includegraphics[width=0.426\textwidth]{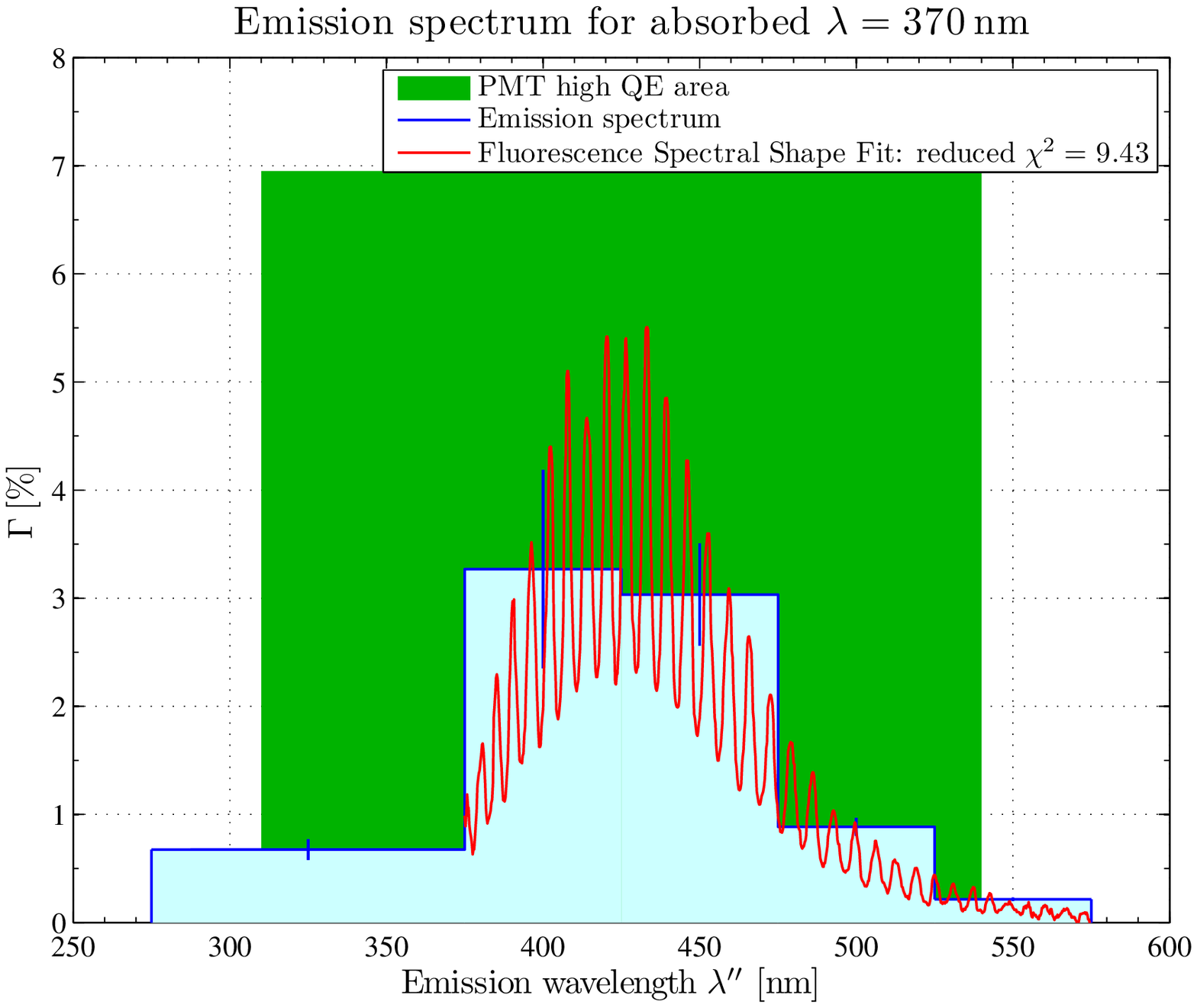}}
 \hfill
 \subfigure[]{\label{fig:WLS:Emissionsspectrum_380nm}\includegraphics[width=0.426\textwidth]{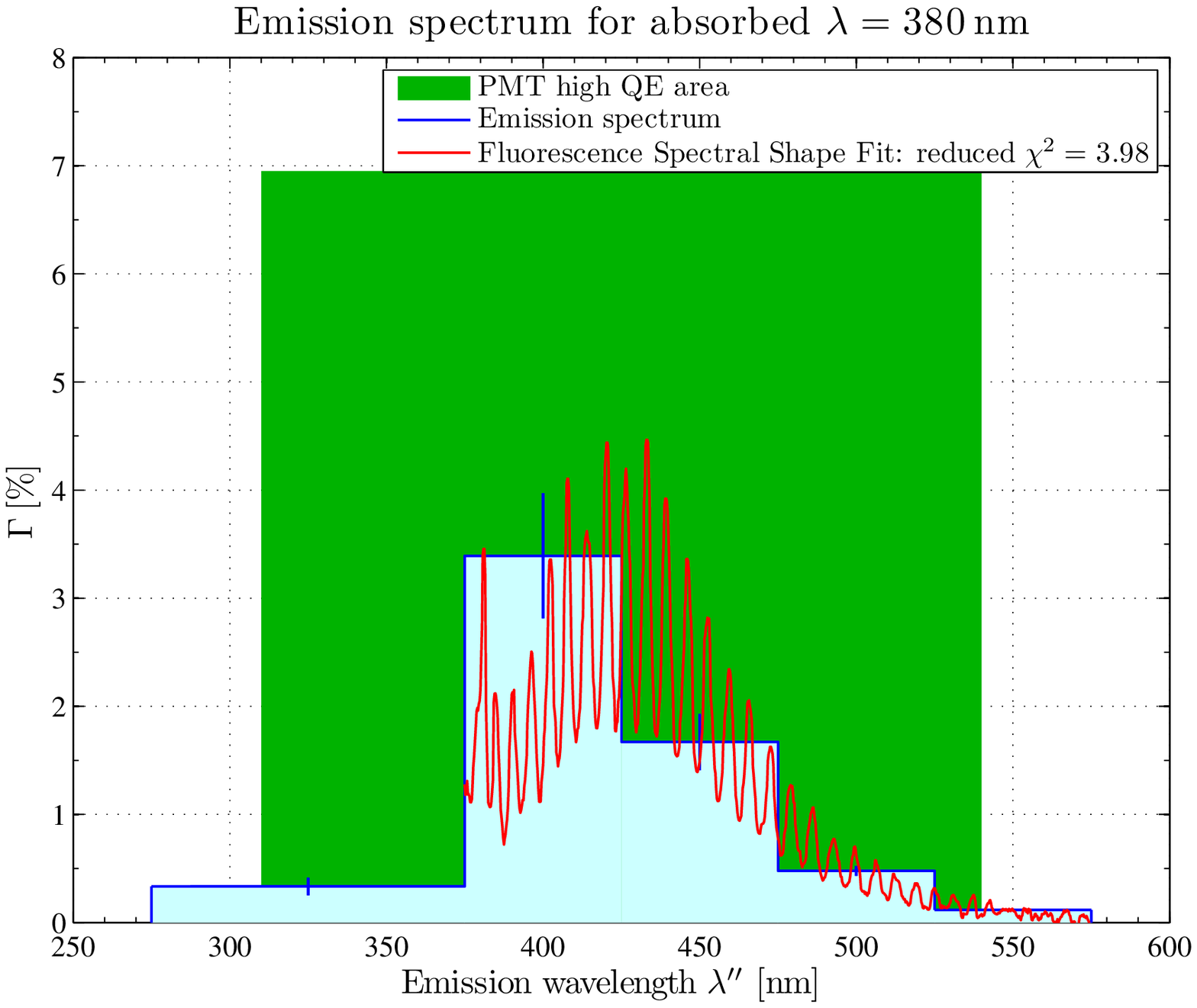}}
 \hfill
 \subfigure[]{\label{fig:WLS:Emissionsspectrum_390nm}\includegraphics[width=0.4265\textwidth]{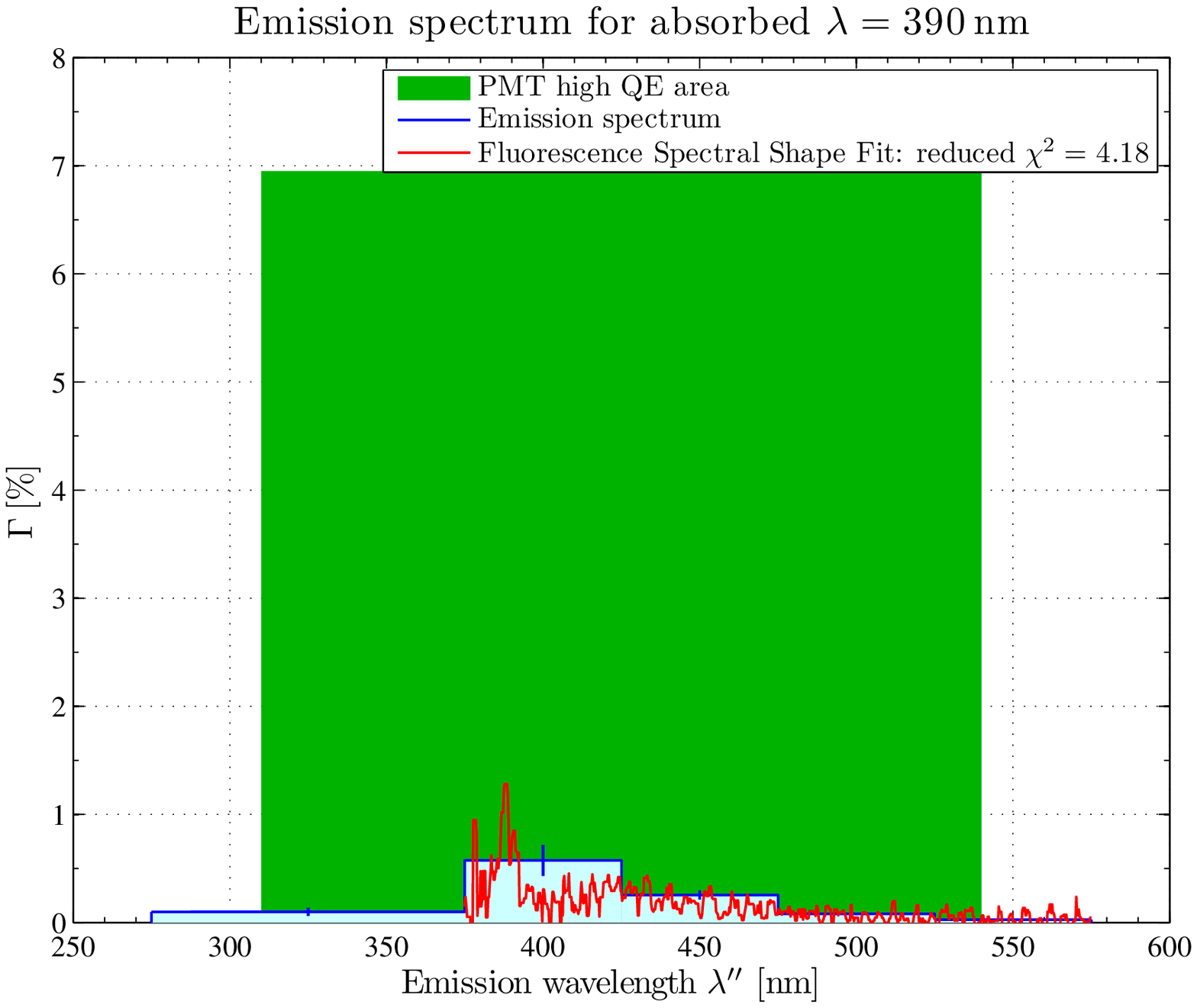}}
 \caption[]{Emission spectra for wavelength shifted light by the DF2000MA foil for different absorption wavelengths \textit{(a)} $\lambda=300\,\mathrm{nm}$, \textit{(b)} $\lambda=320\,\mathrm{nm}$, \mbox{\textit{(c)} $\lambda=370\,\mathrm{nm}$}, \textit{(d)} $\lambda = 380\,\mathrm{nm}$ and \textit{(e)} $\lambda=390\,\mathrm{nm}$ (blue stairs). The WLS spectrum obtained with the spectrometer for each of these wavelengths was scaled by a fit to those values in the trusted wavelength regime of the CCS200 spectrometer (red curve). In figures \textit{(d)} and \textit{(e)} diffuse reflectance is visible as a peak at $\lambda'' = 380\,\mathrm{nm}$ and $\lambda'' = 390\,\mathrm{nm}$, respectively.}
\label{fig:WLS:EmissionSpectra}
\end{figure}

To quantify the wavelength shifting power of the foil, one needs to measure which fraction of the light is shifted to higher wavelength intervals. Such a measurement requires first a wavelength dependent measurement of the intensity of the wavelength shifted light, which was done by measuring the current of the PIN diode $I\left(\lambda,\lambda_{F,i}\right)$ for all filters. The second required information is the total amount of light available for the process of WLS: this information is obtained by measuring the PIN diode current $I_{mf}$ for diffuse reflectance with the reflectance standard described in section~\ref{sec:WLS:setup}. The wavelength dependent ratio of both measurements allows to determine how much light gets wavelength shifted and which wavelength intervals are covered by the emission. Knowing this, it is possible to evaluate the corresponding emission spectrum for every absorption wavelength selected at the monochromator. WLS spectra are visible within the absorbtion wavelength range $250\,\mathrm{nm} \leq \lambda \leq 390\,\mathrm{nm}$. Figure~\ref{fig:WLS:EmissionSpectra} shows exemplarily five different emission spectra for absorption wavelengths $\lambda = 300\,\mathrm{nm}$, $320\,\mathrm{nm}$, $370\,\mathrm{nm}$, $380\,\mathrm{nm}$ and $390\,\mathrm{nm}$. The $y$-axis shows the ratio $\Gamma \coloneqq I\left(\lambda,\lambda_{F,i}\right)/I_{mf} \cdot \rho_{mf}$, while each bin has been corrected for the mean value $R\left(\lambda_{F,i}\right)$ of the PIN diode responsivity within the filter bandwidth (see table~\ref{tab:WLS:Filters}). The highest ratios can be found in figures~\ref{fig:WLS:Emissionsspectrum_300nm}-\subref{fig:WLS:Emissionsspectrum_370nm} ($300\,\mathrm{nm}-370\,\mathrm{nm}$) in a range of $\approx2\%-3\%$. At $\lambda = 380\,\mathrm{nm}$ the diffuse reflectance peak appear at $\lambda'' = 380\,\mathrm{nm}$. This illustrates a systematic uncertainty to the diffuse reflectance evaluation in figure~\ref{fig:WLS:REALDiffuseReflectanceCurve} and has to be corrected for (see section~\ref{sec:systematics:WLS}). For absorption wavelengths of $390\,\mathrm{nm}$ (figure~\ref{fig:WLS:Emissionsspectrum_390nm}), there is already a clearly visible reduction of the WLS, as expected from figures~\ref{fig:refl:AnalysisResultsMainz} and \ref{fig:refl:AnalysisResultsErlangenDifferentPointsCurves}, where the specular reflectance at $390\,\mathrm{nm}$ is almost $100\%$. Besides that, the diffuse reflectance peak can be observed in the PIN diode measurement as well as in the CCS200 spectrometer measurement for that wavelength. No spectrum could be measured above $390\,\mathrm{nm}$. WLS below $250\,\mathrm{nm}$ is probable, but was beyond the measurement range of this setup and could not be verified. As an illustration, the green hatched areas in figure~\ref{fig:WLS:EmissionSpectra} indicate the interval where the muon veto PMTs have a quantum efficiency higher than $10\%$.

Because the enveloping function for the fluorescence emission spectrum of the DF2000MA is unknown, one has to scale the spectrum measured with the CCS200 spectrometer for each wavelength to the bins of the PIN diode measurement. This was done using a one parameter fit $f\left(\lambda''\right)$, scaling the individual spectra best fitting to the PIN diode measurement data for each absorbtion wavelength (shown in red):

\begin{equation}
	f\left(\lambda''\right) = a_0 \cdot \varphi\left(\lambda''\right), \label{eqn:WLS:Fit}
\end{equation}

with $\varphi\left(\lambda''\right)$ being the emission spectrum obtained by the CCS200 spectrometer individually for each absorption wavelength $\lambda$. The fit parameter $a_0$ is found to scale the measured spectral shape best to the PIN diode measurement data. As stated in section~\ref{sec:WLS:setup}, the spectrometer can only be used for emission wavelength $\geq 375\,\mathrm{nm}$, so the fit was performed in the interval $375\,\mathrm{nm} \leq \lambda \leq 575\,\mathrm{nm}$, where $575\,\mathrm{nm}$ is the last wavelength included in the bandwidths of one of the optical filters. The shape of the spectrum suggests that the true emission curve also extends to values below $375\,\mathrm{nm}$ (especially in figure~\ref{fig:WLS:Emissionsspectrum_370nm}), but only the PIN diode is sensitive to this part of the WLS emission within the $100\,\mathrm{nm}$ wide bandwidth of filter 1. The single peaks of the vibrational sub energy niveaus are clearly visible and have a fixed separation of $6\,\mathrm{nm}$. The highest emission values of the WLS spectrum can be observed at $\approx420\,\mathrm{nm}$. The typical width of the distribution at the half of its maximum is $\approx80\,\mathrm{nm}$ (figure~\ref{fig:WLS:EmissionSpectra}). Those seperations are independent of the absorption wavelength and have been used as an input for the XENON1T MC simulation (see section~\ref{sec:MC}).

\begin{figure}[htb]
 \centering
 \subfigure[]{\label{fig:WLS:Absorptionspectrum_325-375nm}\includegraphics[width=0.45\textwidth]{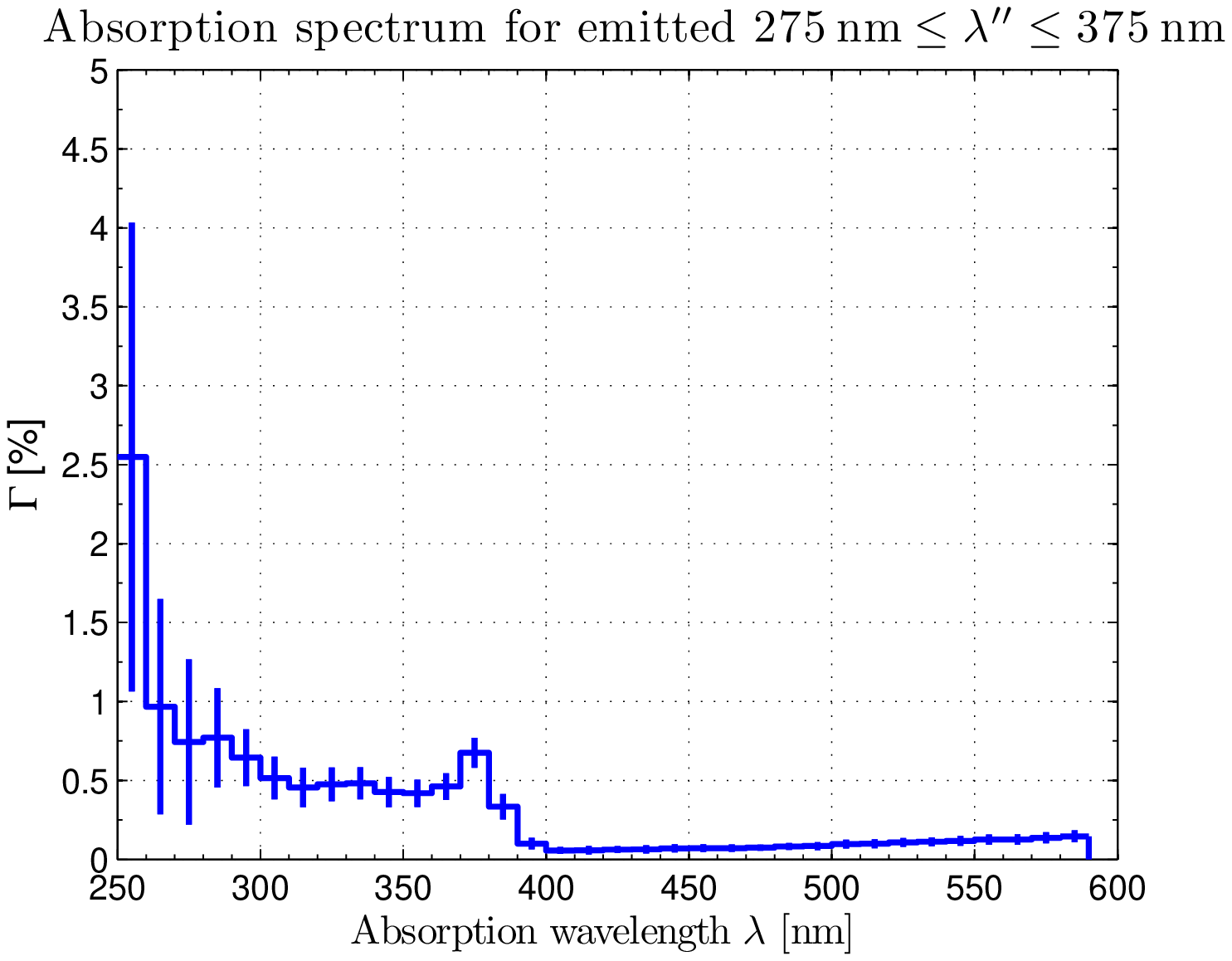}} 
 \hfill
 \subfigure[]{\label{fig:WLS:Absorptionspectrum_375-425nm}\includegraphics[width=0.45\textwidth]{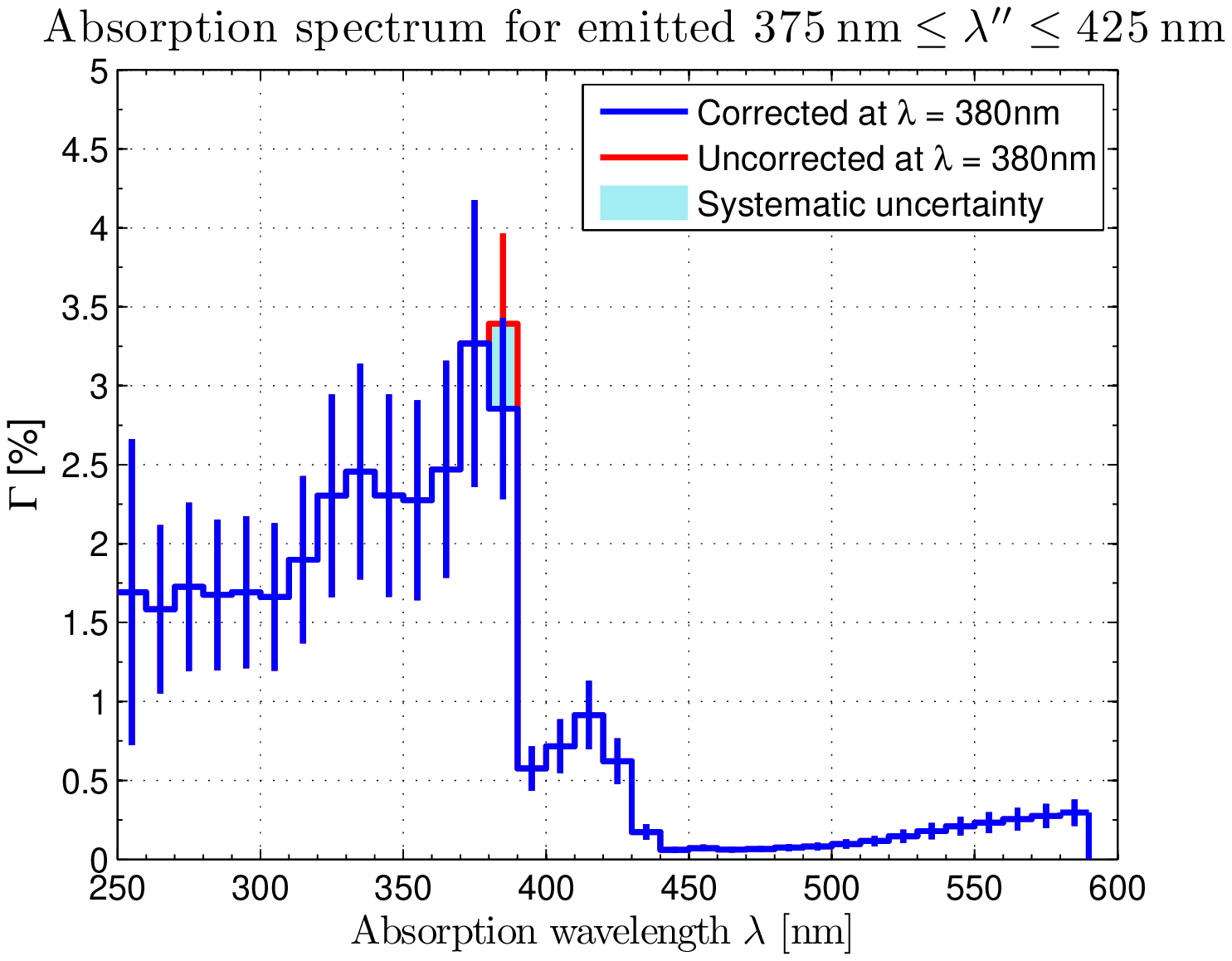}}
 \hfill
 \subfigure[]{\label{fig:WLS:Absorptionspectrum_425-475nm}\includegraphics[width=0.45\textwidth]{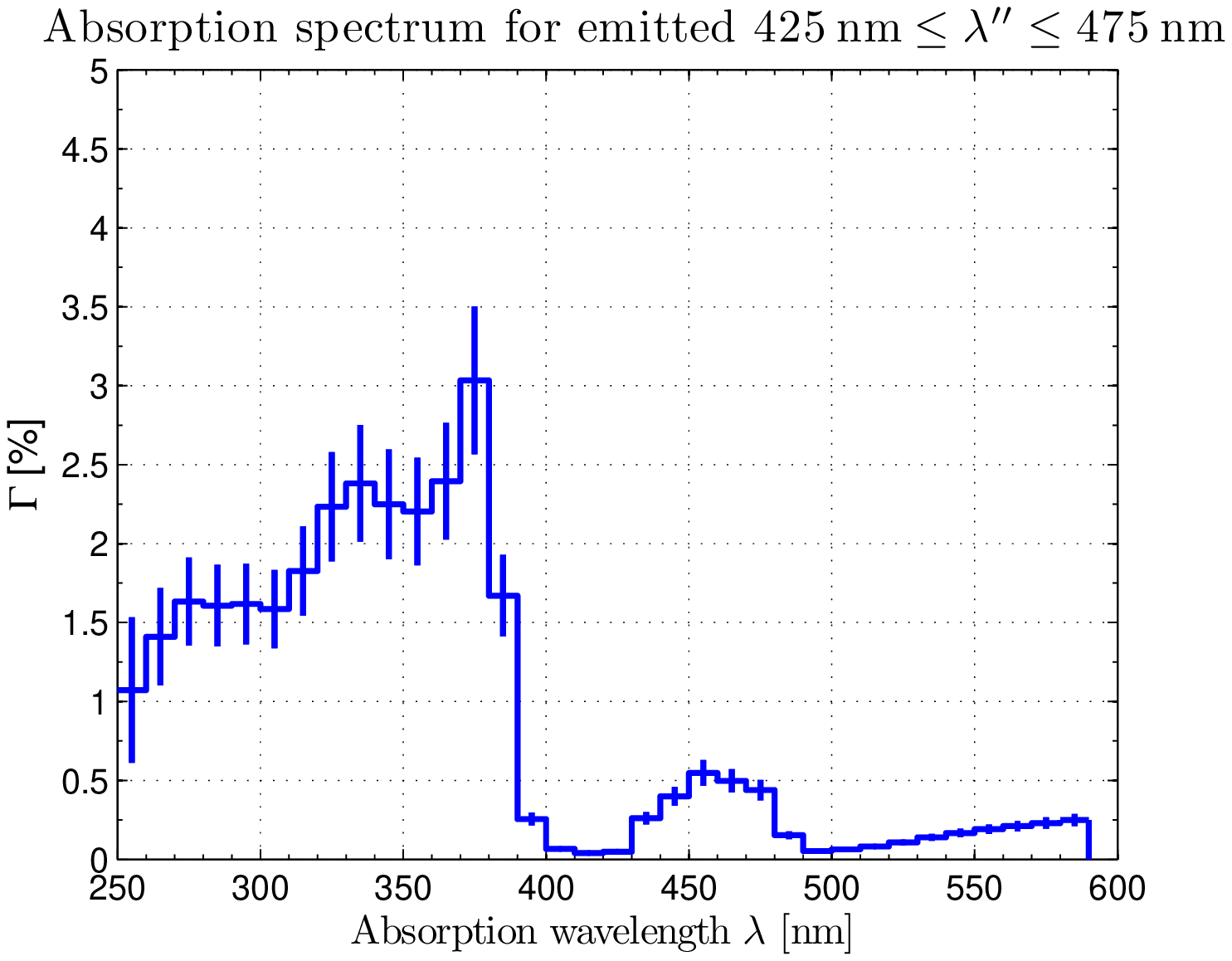}}
 \hfill
 \subfigure[]{\label{fig:WLS:Absorptionspectrum_475-525nm}\includegraphics[width=0.45\textwidth]{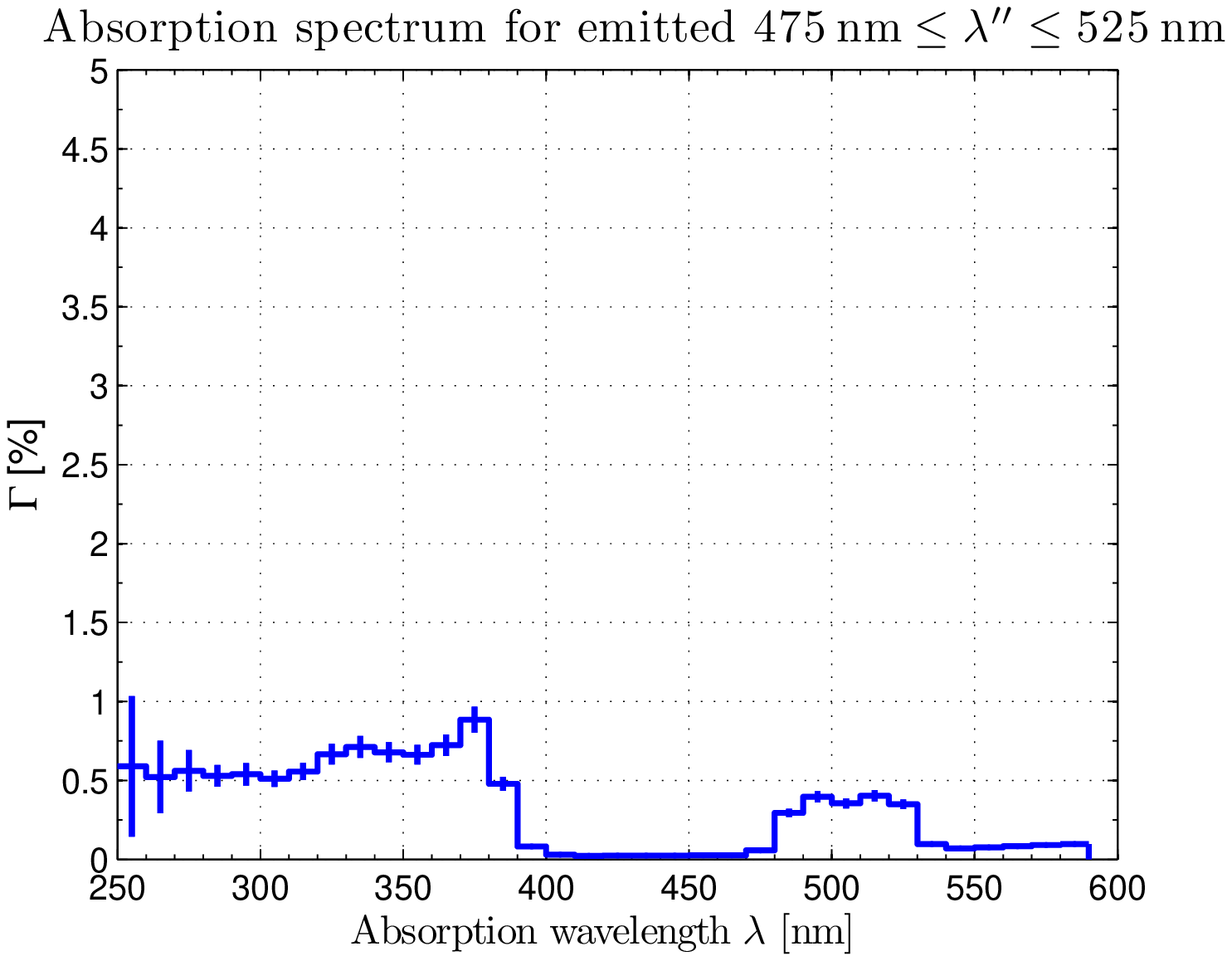}}
 \hfill
 \subfigure[]{\label{fig:WLS:Absorptionspectrum_525-575nm}\includegraphics[width=0.45\textwidth]{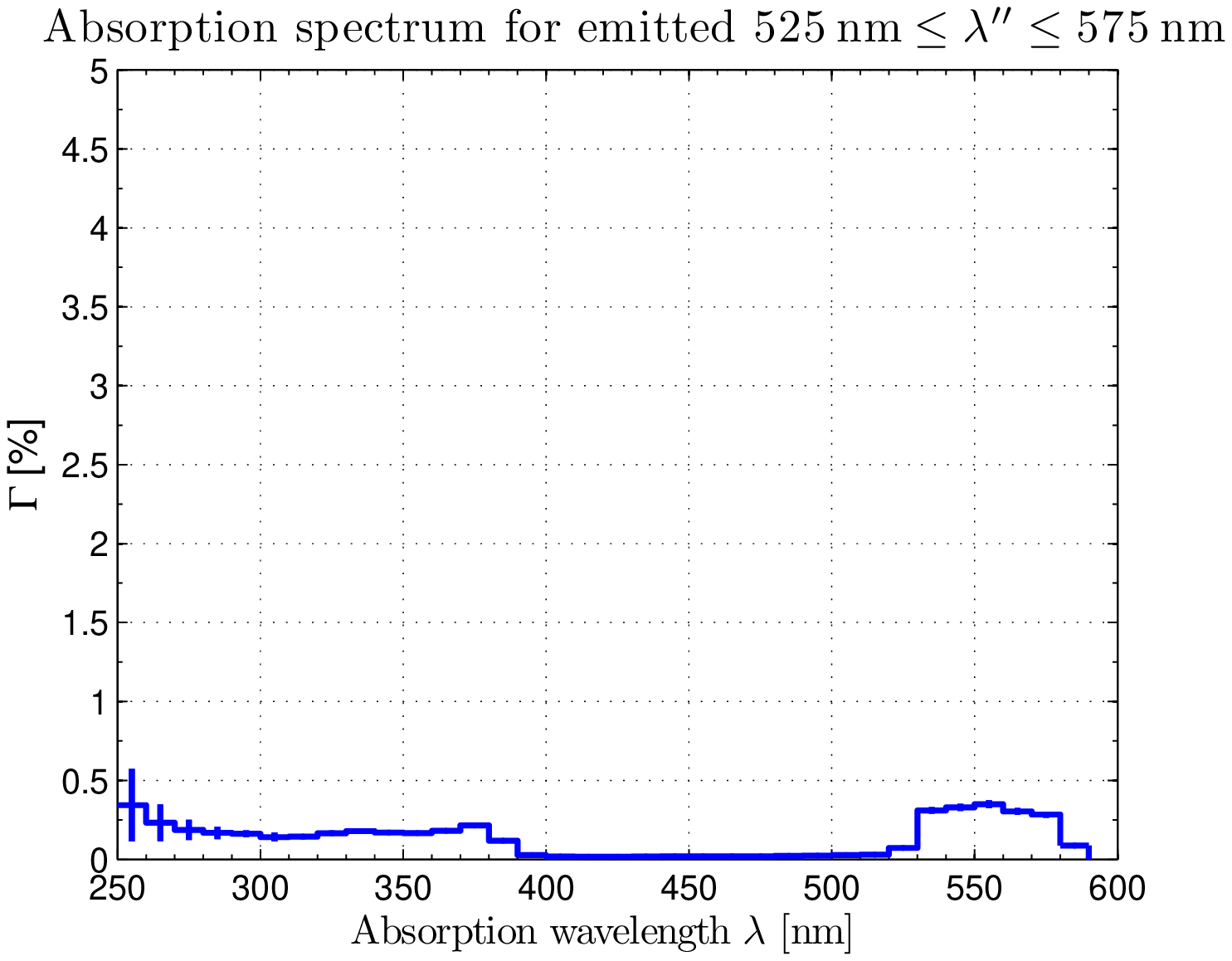}}
 \caption[]{Absorption spectra for each emission interval of wavelength shifted light by the DF2000MA foil according to the five different optical bandwidth filters. \mbox{\textit{(a)} Filter 1: $275\,\mathrm{nm}-375\,\mathrm{nm}$}; \mbox{\textit{(b)} Filter 2: $375\,\mathrm{nm}-425\,\mathrm{nm}$}, where a correction for the contribution of diffuse reflectance at \mbox{$\lambda = 380\,\mathrm{nm}$} was applied, lowering the corresponding data point; \mbox{\textit{(c)} Filter 3: $425\,\mathrm{nm}-475\,\mathrm{nm}$}; \textit{(d)} \mbox{Filter 4: $475\,\mathrm{nm}-525\,\mathrm{nm}$}; \textit{(e)} \mbox{Filter 5: $525\,\mathrm{nm}-575\,\mathrm{nm}$}.}
\label{fig:WLS:AbsorptionSpectra}
\end{figure}

Furthermore, the measurement method makes it possible to plot an absorption spectrum for each optical filter bandwith used during the measurement. This allows to determine the absorbed wavelength for every emission wavelength interval. The five plots for the five different emission wavelength ranges are shown in figure~\ref{fig:WLS:AbsorptionSpectra}. 

The $x$ axis shows the absorbed wavelength $\lambda$, while the $y$ axis again stands for the ratio \mbox{$\Gamma = I\left(\lambda,\lambda_{F,i}\right)/I_{mf}\cdot \rho_{mf}$}. The results confirm the curves shown in figure~\ref{fig:WLS:EmissionSpectra}, since here one can find the highest value of the light ratio to be in the emission interval $375\,\mathrm{nm} \leq \lambda'' \leq 475\,\mathrm{nm}$ (figures~\ref{fig:WLS:Absorptionspectrum_375-425nm} and \subref{fig:WLS:Absorptionspectrum_425-475nm}). It turns out that the DF2000MA foil absorbs light up to wavelengths smaller than $400\,\mathrm{nm}$, where the high specular reflectance starts. In all figures, one can see the effect of the diffuse reflectance as peaks located at the emission intervals. In figure~\ref{fig:WLS:Absorptionspectrum_375-425nm} a correction was applied for the data point of $\lambda = 380\,\mathrm{nm}$. It is a vice versa correction compared to figure~\ref{fig:WLS:REALDiffuseReflectanceCurve} as here the contribution of diffuse reflectance have to be corrected for. The non-zero values at large wavelengths outside the emission intervals are likely caused by scattered light on the shiny optical table surface inside the dark box.

\begin{figure}[htb]
	\centering
	\includegraphics[width=0.70\textwidth]{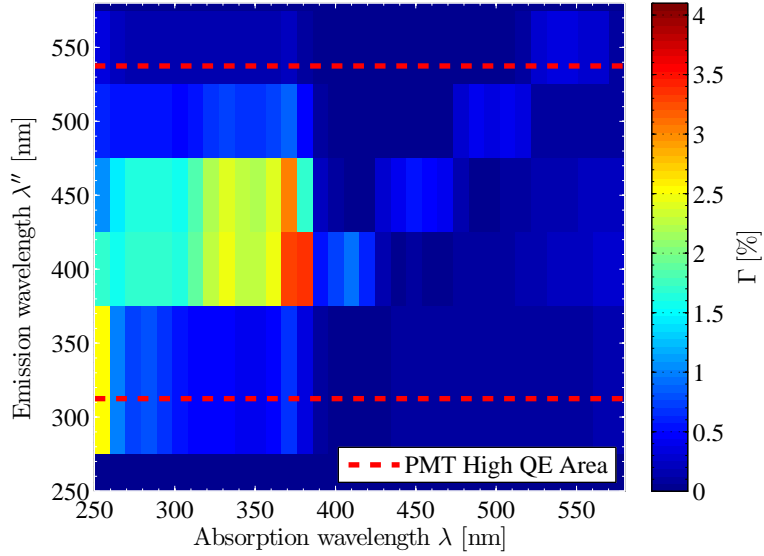}
	\caption{Diffuse response of the DF2000MA foil including wavelength shift and diffuse reflection. The WLS emission maximum is between \mbox{$\lambda'' = 375\,\mathrm{nm}$} and $475\,\mathrm{nm}$ and lies in between the high QE area of the muon veto PMTs (red dashed lines). Highest values for the ratio of WLS light to entire light can be found up to $3.5\%$. The diffuse reflectance peaks are visible as small values along the diagonal for $\lambda = \lambda''$, overlapping with the WLS spectrum at $\lambda = 380\,\mathrm{nm}$.}
	\label{fig:WLS:Colormap}
\end{figure}

Merging the results of figures~\ref{fig:WLS:EmissionSpectra} and \ref{fig:WLS:AbsorptionSpectra} leads to a two dimensional histogram shown in figure~\ref{fig:WLS:Colormap}. As warmer the color gets, the higher the percentage of the light ratio is. The single spectra of figure~\ref{fig:WLS:EmissionSpectra} correspond to vertical lines in the histogram, while the spectra of figure~\ref{fig:WLS:AbsorptionSpectra} correspond to horizontal lines. 

\begin{figure}[htb]
	\centering
	\includegraphics[width=0.70\textwidth]{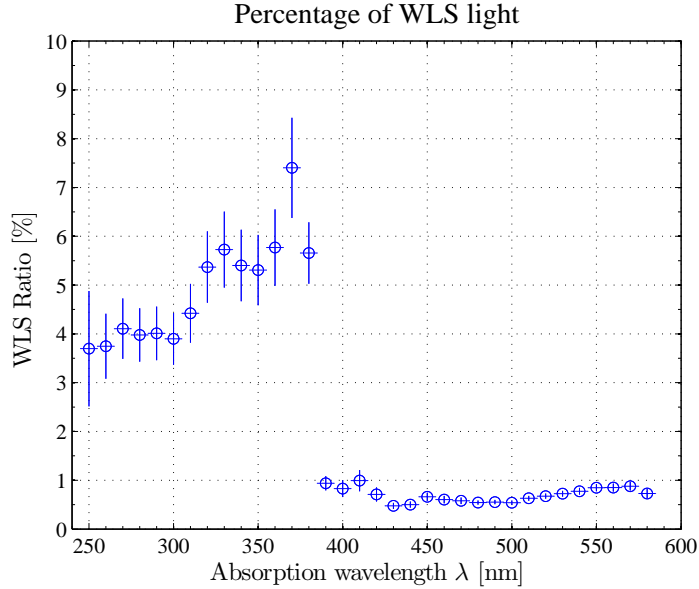}
	\caption{Percental amount of WLS light compared to the entire amount of light reflected by the DF2000MA foil for all emission wavelengths.}
	\label{fig:WLS:Percentage}
\end{figure}

Almost all absorption takes place for wavelengths lower than $400\,$nm. The emission spectrum reaches from $375$ to $575$ nm and covers all bandwidths of the used optical filters. Additional WLS to higher wavelength intervals ($\lambda > 575\,\mathrm{nm}$) was not observed during the measurements with the spectrometer, but would be anyway out of the range of interest defined by the high QE areas of the XENON1T muon veto PMTs. Once every emission spectrum is scaled with a spectral shape distribution, the spectrum values are integrated over the emission wavelength $\lambda''$ to obtain the amount of fluorescence WLS light $\sum_i I\left(\lambda,\lambda_{F,i}\right)$ relative to the entire amount of light $I_{mf}/\rho_{mf}$, which is just dependent on the absorption wavelength $\lambda$, but not on the emitted wavelength $\lambda''$. This quantity is called WLS ratio. Figure~\ref{fig:WLS:Percentage} shows the WLS ratio as extracted from the measurement. The DF2000MA foil absorbs most of the light at the wavelength of $370\,\mathrm{nm}$ where the highest percentages of WLS light can be found. The maximum relative value for the amount of WLS light compared to the entire light amount is $\approx7.5\%$ for an absorption wavelength of $370\,$nm. This means that out of all photons exciting electrons to a higher energetic state by absorption in the foils molecules, just those $7.5\%$ get reemitted in form of fluorescence light. The energy of the residual absorbed light is released via radiationless heat transfer, not measurable by this experimental setup. The shape of the curve verifies also the results described in section~\ref{sec:refl:analysis}, since it is roughly inverse to the reflectance curve of figure~\ref{fig:refl:AnalysisResultsMainz}. 

\begin{figure}[htb]
	\centering
	\includegraphics[width=0.70\textwidth]{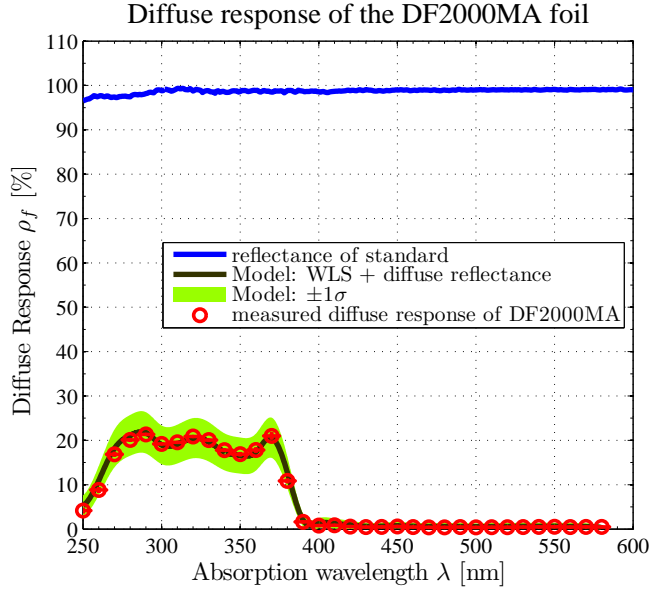}
	\caption{Response of the DF2000MA foil to a filter-less setup. The three peaks are an indication of absorption maxima since light gets absorbed at these wavelengths and is shifted to larger wavelengths. The components of diffuse reflectance and wavelength shifting are superpositioned in this plot (dark-green line), but not distinguishable in the filter-less measurement (red circles).}
	\label{fig:WLS:DiffuseReflectanceCurve}
\end{figure}

A model of the diffuse response $\rho_f$ from \eqref{eqn:WLS:DiffuseResponse} can be built with the light amount used for wavelength shifting and the one used for diffuse reflectance. To cross-check this, the model was compared to a separate measurement with the diffuse setup without the usage of any filter. This yields the curves shown in figure~\ref{fig:WLS:DiffuseReflectanceCurve}. In blue, the reflectance curve of the diffuse reflectance standard is shown as a function of the wavelength $\lambda$ selected with the monochromator, while the dark green line represent the model (wavelength shifting + diffuse reflectance) with the uncertainty bands of $\pm 1\,\sigma$ indicated in light green. The red circles display the response $\rho_f$ of the DF2000MA foil obtained by a separate measurement without filters and with an uncertainty of $\mathcal{O}\left(1\%\right)$. Model and data are in good agreement. $\rho_f$ is smaller than $0.5\%$ for $\lambda > 400\,\mathrm{nm}$. One can see that light below $400\,\mathrm{nm}$ is absorbed and can be used for WLS.

\section{Monte Carlo Simulation}\label{sec:MC}

The XENON1T muon veto system was designed and based on a dedicated MC simulation described in \cite{Fattori_MV}. The study was carried out with the Geant4 toolkit \cite{Geant4}. The geometry of the simulation included the DF2000MA foil on the inner water tank surface with preliminary reflectance values for different wavelengths, based on the values provided by the manufacturer and preliminary results of this study. The WLS power of the foil was not included. The paper presents the detection efficiency of the muon veto for two different physical scenarios:

\begin{enumerate}                                                                                                                                                                                                                                                                                                                                                                                                                                                                                                                                                                                
  \item [i)] The parent muon enter the water tank, which is called ``muon event'' and happens in $\approx1/3$ of all cases;
  \item [ii)] The parent muon passes the outside and just its secondary particles enter the water tank, which is called ``shower event'' and happens in $\approx2/3$ of all cases;
\end{enumerate}

\begin{figure}[htb]
  \centering
    \subfigure[]{\label{fig:WLS:MC:ModRefl}\includegraphics[width=0.45\textwidth]{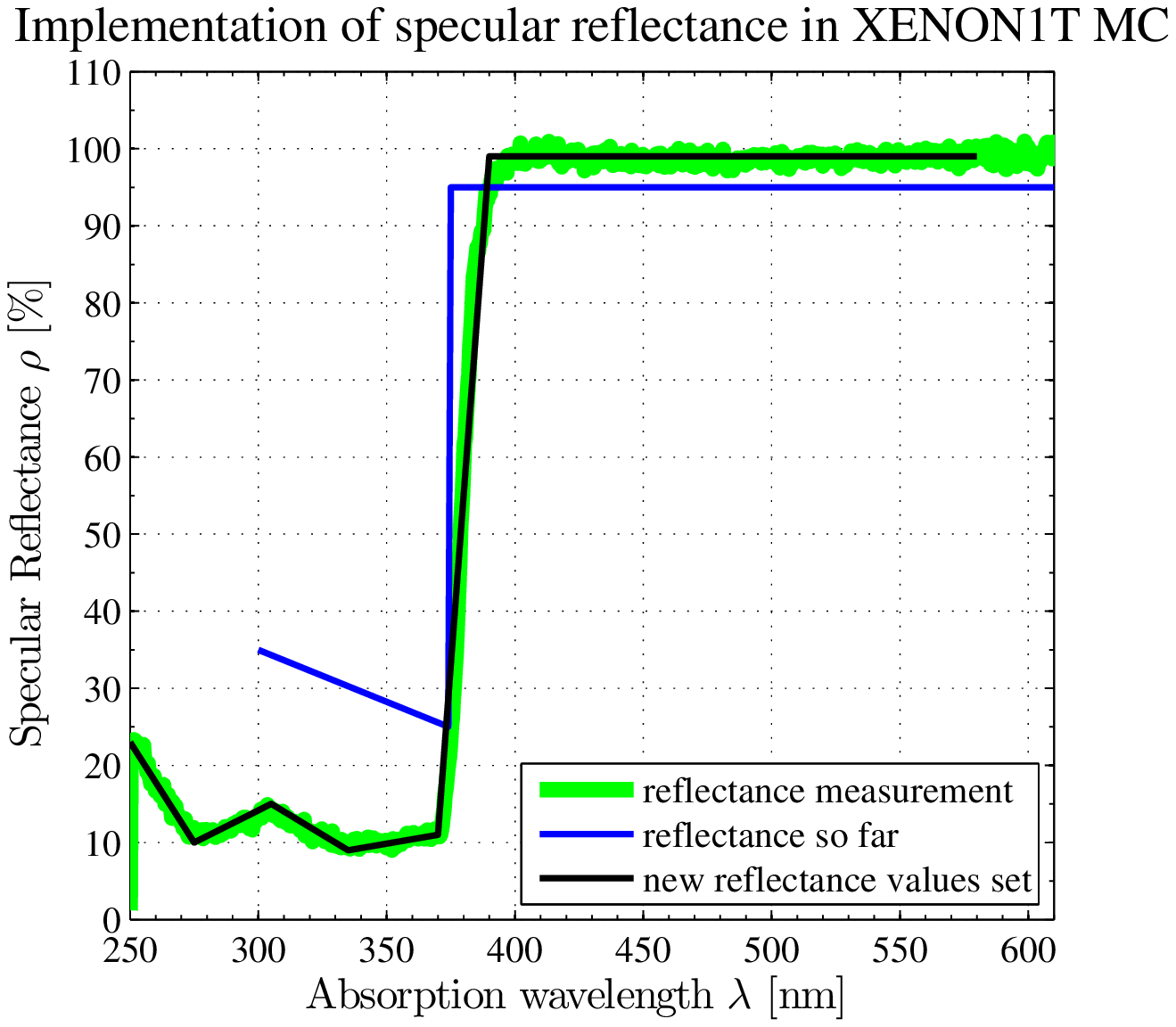}}
    \subfigure[]{\label{fig:WLS:MC:RecalcColormap}\includegraphics[width=0.45\textwidth]{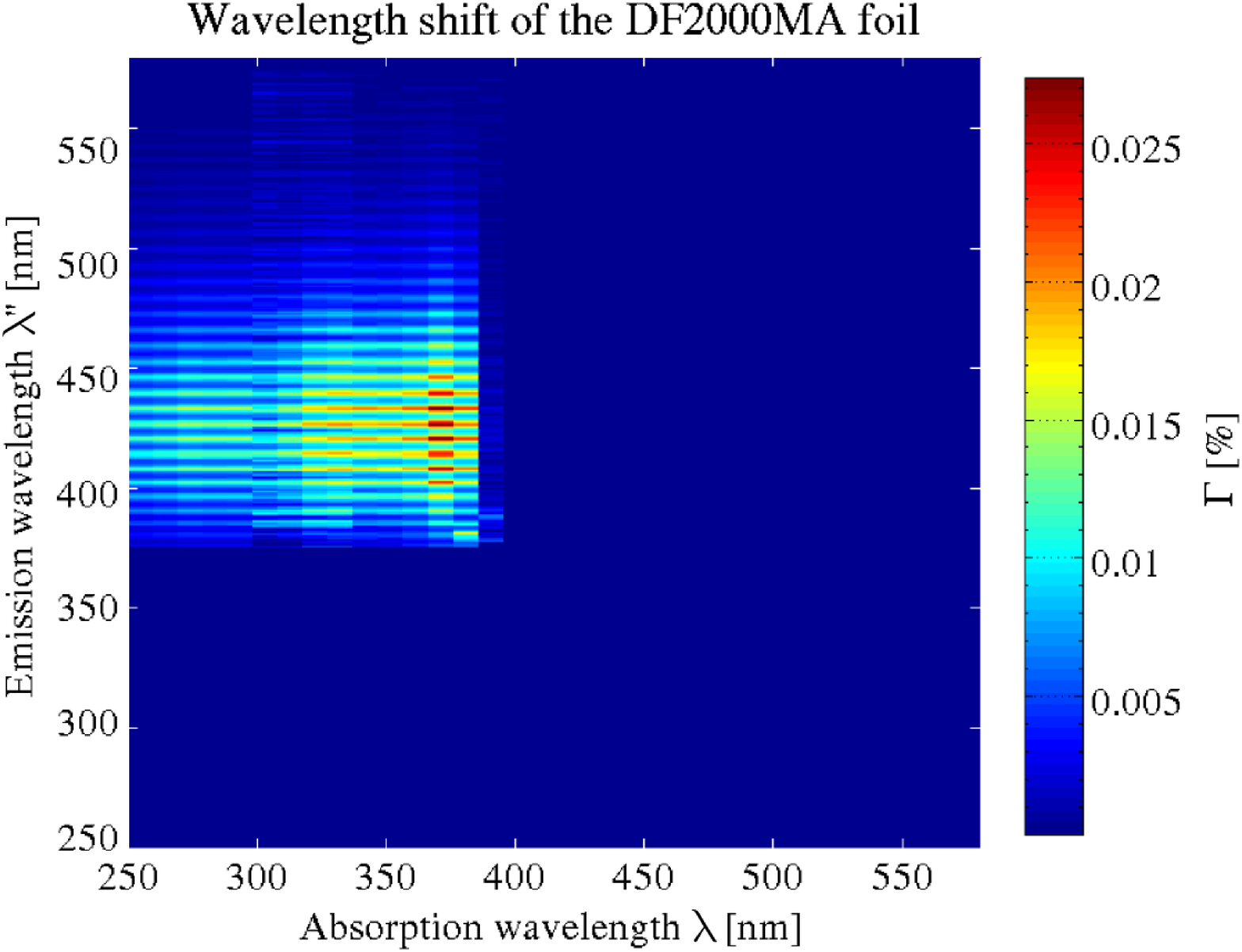}}
    \caption[]{{\em a)} Different implementations of the DF2000MA reflectance in the Geant4 simulation. The blue curve shows the reflectance as it was defined in the MC study of \cite{Fattori_MV}, the black curve shows the reflectance as it was defined in this work. The green curve is the result of the reflectance measurement; {\em b)} WLS absorption versus emission wavelength. The resolution on the emission side was improved by fitting the emission spectra for each absorption wavelength with the spectral shape of the fluorescence spectrum obtained by the CCS200 spectrometer.}
  \label{fig:WLS:MC:MeasurementImplementation}
\end{figure}

The study provides an efficiency of $\left(99.78\pm0.05\right)\%$ for the ``muon event'' case and $\left(70.6\pm0.5\right)\%$ for the ``shower event'' case. Here, we present the results of the same MC simulation, but with modified reflectance properties of the DF2000MA foil according to the values presented in this work (compare figure~\ref{fig:WLS:MC:ModRefl}) and after implementation of the WLS power of the foil.

To implement the WLS in the Geant4 simulation, the resolution of the emission spectra (compare figures~{\ref{fig:WLS:EmissionSpectra}) has to be improved. The resolution of the emission spectra is limited by the bandwidth of the used optical filters. Fitting the fluorescence spectral shape obtained by the CCS200 spectrometer for each absorption wavelength $\lambda$ to the corresponding emission spectrum (as described in section~\ref{sec:WLS:analysis}) allows a finer binning of the latter, limited only by the resolution of the spectrometer (i.e. $0.25\,\mathrm{nm}$). Figure~\ref{fig:WLS:MC:RecalcColormap} shows the 2D histogram shown in figure~\ref{fig:WLS:Colormap} with the adjusted resolution for the emission wavelength, based on the information obtained with the spectrometer measurement. 

Running the simulations as in \cite{Fattori_MV} with the modified reflectance values in the Geant4 simulation, the detection efficiency of the muon veto remains almost unchanged at $\left(99.72\pm0.06\right)\%$ for ''muon events" as well as for ''shower events'' with $\left(70.9\pm0.5\right)\%$. The effects of the overestimated reflectance below $370\,\mathrm{nm}$ in the original MC simulation (shown in the blue curve in figure~\ref{fig:WLS:MC:ModRefl}) and the additional implemented data points below $300\,\mathrm{nm}$ compensate each other. If, in addition, wavelength shifting is allowed, the detection efficiency for ``muon events'' remains again almost unchanged with $\left(99.69\pm0.06\right)\%$ but rises $\approx0.5\%$ for shower events to $\left(71.1\pm0.5\right)\%$. This is, within the uncertainty of the simulation, a significant but small increase of the muon veto efficiency due to the wavelength shifting of the DF2000MA foil. This is consistent with the physics in the muon veto. The amount of Cherenkov light created by muons is so big ($\approx3000$\,photons per muon), that the WLS effect has no impact, while for the shower case, due to the lower number of produced photons ($\approx400$\,photons per shower), one could expect a small effect on the efficiency.

\section{Assessment of Systematic Uncertainties}
\label{sec:systematics}

\subsection{General Systematics}
\label{sec:systematics:general}

Several uncertainties have to be assessed for the measurement method presented here. The Xe light source shows a drift in intensity during a heat-up phase. To ensure stable light source conditions, measurements started not earlier than $30$\,minutes after switching on the lamp. After heat-up the lamp is cooled by a controlled fan unit to maintain optimal operation temperature. The measurement setups were located inside a walkable laminar flow box with a constant air flow from top. Humidity absorption into the sensitive area of the PIN diode can be excluded therefore. Systematics induced by dirt on any of the optical components was avoided by wearing fabric gloves. All measurements were performed within one year. Thus, aging effects of the PIN diode should not induce any systematics. The laminar flow box was temperature controlled to $T = \left(294.0 \pm 0.3\right)\,$K. Applying the Shockley diode equation, the expected fluctuations of the temperature dependent PIN diode dark current are $\approx \pm 4\,\%$. Sporadic dark current measurements showed values $\mathcal{O}\left(0.1\,\mathrm{pA}\right)$, while the smallest photocurrent measurements were $\mathcal{O}\left(0.1\,\mathrm{nA}\right)$. The dark current fluctuation by temperature changes is therefore negligible. The same holds for the quantum efficiency of the PIN diode and its influence on the responsivity $R$ (see equation \eqref{eqn:refl:Responsivity}). Given the temperature dependence on the band gap of silicon one can calculate an effect of $\pm 10^{-5}$, which is also negligible.

\subsection{Systematics of the Specular Reflectance Measurements}
\label{sec:systematics:refl}

All geometrical systematics like distances, solid angles, angles of incidence, etc. remained constant during the measurements and cancel out after applying equation \eqref{eqn:refl:SpecularReflectance}. This is enabled by the relative measurement of data values obtained with the DF2000MA and the reflectance standard with the same setup, as described in section~\ref{sec:refl:setup}. The same holds for induced systematics by the lens transfer function $L\left(\lambda\right)$. The spectral shape of the light exiting the monochromator is triangular with FWHM of $G = 0.38\,\mathrm{nm}$ for a slit size $s = 0.07\,\mathrm{mm}$, according to specifications in the manual. As this is much smaller than the resolution of the remaining setup, this is approximated by a $\delta$ function as described in section~\ref{sec:refl:analysis}.

\subsection{Systematics of the Diffuse Reflectance and WLS Measurements}
\label{sec:systematics:WLS}

Geometrically induced systematics are eliminated due to the relative definition of $\Gamma$ in section~\ref{sec:WLS:analysis}, similar to the reflectivity measurements. Contrary to the specular reflectance measurements the spectral shape of the monochromator output light has for the widely opened slit ($s = 2\,$mm) a rectangular shape with FWHM of $G = 10.81\,\mathrm{nm}$. However, this bandwidth is in accordance with the chosen wavelength steps selected at the monochromator and is small compared to the used filter bandwidths as described in section~\ref{sec:WLS:analysis}. Here, the transmission functions of the filters do not cancel out and the data values had to be corrected for their corresponding transmittance $\tau$ (see table~\ref{tab:WLS:Filters} and text description). Further, a careful test was performed in order to optimize the distance $x$ between filter and foil surface. It is not allowed to be too small since ghost reflections between the foil surface and the surface of the filters could occur. This happens because the transmission of the filters is not $100\%$ and light can be trapped by multiple reflections between filters and foil sample. This reduces the measurable intensity but avoids systematic uncertainties from ghost reflections.

For the absorption wavelength of $\lambda = 380\,$nm diffuse reflectance and WLS emission add to each other, leading to a systematic uncertainty. Its effect on the diffuse reflection and WLS analysis can be estimated: the clearly visible diffuse reflectance peak at the beginning of the filter interval $375\,\mathrm{nm} \leq \lambda_{F,2} \leq 425\,\mathrm{nm}$ in figure~\ref{fig:WLS:Emissionsspectrum_380nm} indicates the systematic of diffuse reflectance and WLS being present within one filter bandwidth. Thus, there is a good chance that light of $\lambda = 380\,\mathrm{nm}$ gets partly diffusively reflected at $\lambda'' = 380\,\mathrm{nm}$ and partly gets absorbed and re-emitted wavelength-shifted at e.g. $\lambda'' = 420\,\mathrm{nm}$. Because both wavelengths are within the same filter bandwidth, the analysis method cannot distinguish between diffuse reflectance and WLS. WLS light contributes to the calculated diffuse reflectance value and systematically shifts it to higher values (red circle in figure~\ref{fig:WLS:REALDiffuseReflectanceCurve}) as well as, vice versa, diffuse reflectance contributes to the WLS emission (red histogram in figure~\ref{fig:WLS:Absorptionspectrum_375-425nm}). Excluding the absorption wavelength $\lambda = 380\,$nm from diffuse reflectance analysis leads to the blue circle in figure~\ref{fig:WLS:REALDiffuseReflectanceCurve} and the blue histogram in figure~\ref{fig:WLS:Absorptionspectrum_325-375nm}. The real value in both plots is somewhere in between (indicated by the light blue error band), but is not determinable given the $50\,\mathrm{nm}$ resolution of the used measurement method. For diffuse reflectance the hereby induced systematic error for this data point is $\approx 75\%$, for WLS emission it is $\approx 15\%$. Absorption wavelengths of $370\,\mathrm{nm}$ or $390\,\mathrm{nm}$ do not induce this systematic, since $370\,\mathrm{nm}$ is at the end of a filter interval (so WLS light is not transmitted), while for $\lambda = 390\,\mathrm{nm}$ the foils specular reflectance is already high enough to almost not allow for absorption and WLS anymore (see figures~\ref{fig:WLS:Emissionsspectrum_370nm} and \ref{fig:WLS:Emissionsspectrum_390nm}).

\section{Conclusions} \label{sec:concl}

The DF2000MA foil optical response was inspected in detail. Figure~\ref{fig:concl:results} shows the results summarized in one plot. In case of the specular reflectance, almost $100\%$ of the incoming light gets reflected for all wavelength larger than $400\,\mathrm{nm}$. The rise of the reflectance starts already at $370\,\mathrm{nm}$. These values are $30\,\mathrm{nm}$ below the values provided by the manufacturer in 2007 \cite{3M} and are consistent with new values provided by 3M in 2017 \cite{3M_2017}. For wavelengths $\lambda > 450\,\mathrm{nm}$ no difference in reflectance has been observed for the two angles of incidence tested. Below $370\,\mathrm{nm}$, deviations of $\approx 2\%$ in reflectance have been measured. In addition, the high reflectance value is independent of the position on the foil area. For wavelength smaller than $370\,\mathrm{nm}$ just $\approx10\%$ of the light gets reflected with fluctuations of about $\pm\,5\%$, depending on the wavelength. The remaining amount of light gets absorbed by the foil. Therefore, the high reflectance range covers $\approx74\%$ of the high sensitive wavelength range of the muon veto PMTs ($310\,\mathrm{nm}-540\,\mathrm{nm}$).
Diffuse reflectance was observed within a $0.3\%-0.7\%$ level and is negligible within the context of the foil use in the XENON1T experiment. The light absorbed by the foil is largely transfered into heat. Just $\approx3-7.5\%$ of the incoming light is used for WLS, depending on the incoming wavelength. 

\begin{figure}[htb!]
	\centering
	\includegraphics[width=0.70\textwidth]{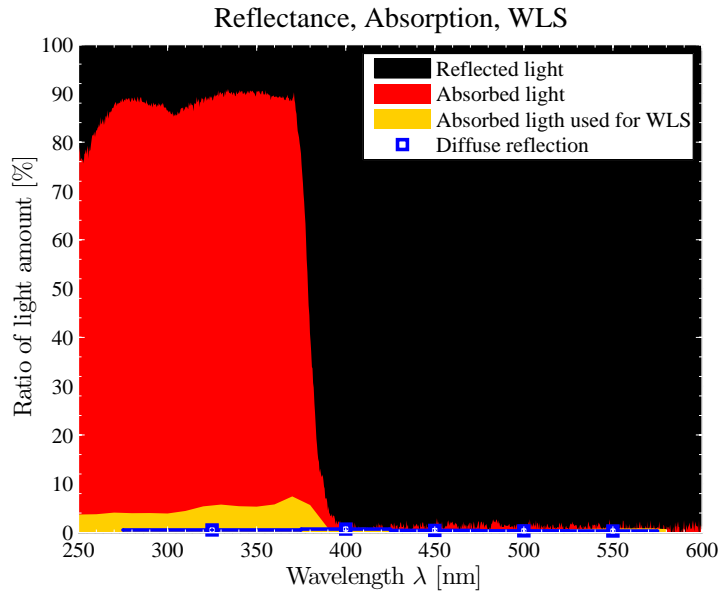}
	\caption{Pro rata visualization of how much light gets reflected, absorbed and wavelength shifted as a function of the absorption wavelength $\lambda$. About $90\%$ of the light with a wavelength lower than $370\,\mathrm{nm}$ gets absorbed, while up to $7.5\%$ gets later re-emitted as WLS light. Approximately $10\%$ gets reflected specularily. For wavelength greater than $400\,\mathrm{nm}$ there is no absorption anymore. The diffuse reflection (blue squares) is less than $1\%$.}
	\label{fig:concl:results}
\end{figure}

The foil absorbs light of all inspected wavelengths below $400\,\mathrm{nm}$, with $250\,\mathrm{nm}$ being the lowest one. The emission spectrum has the typical shape of a rotational-vibratonal spectrum of fluorescence with a separation between the single transition peaks of $6\,\mathrm{nm}$. The maximum of the fluorescence light is at an emission wavelength of $\approx420\,\mathrm{nm}$. The mean width of the spectrum is $80\,\mathrm{nm}$. For absorption wavelengths greater than $400\,\mathrm{nm}$ no WLS is observed. The influence of superficial destructions and damage of the foil on the reflectance is strongly dependent on the severity of the damage. Slight scratches, like they could happen during the attachment of the foil on some surface, lower the reflectance up to $5\%$ independently from the wavelength. Severe damage and deep scratches on the foils surface lower the reflectance for wavelengths greater than $400\,$nm by up to $40\%$.
Implementation of the more accurate reflectance measurement results into the muon veto Monte Carlo simulation leads to no significant changes of the detection efficiency for muon events and shower events with respect to the previous study \cite{Fattori_MV}. Implementing the WLS proccess to the simulation shows a relative increase of $\approx0.5\%$ only for the shower events, which represents, within the statistical error ranges of the simulation, no significant gain of the muon veto efficiency.

% \appendix
% \section{Some title}
% Please always give a title also for appendices.

\acknowledgments

We gratefully acknowledge support from Dr. Friedrich Stinzing of the Erlangen Center for Astroparticle Physics (ECAP) for providing the opportunity to use their Ocean Optics JAZ spectrometer. We are furthermore grateful to Dr. Marcus Beck of the Johannes Gutenberg University of Mainz for providing a different monochromator and the lab space in a preliminary setup of this study.


\end{document}